\documentclass[twocolumn,superscriptaddress,aps,pra,preprintnumbers,showkeys,10pt]{revtex4-2}
\usepackage{graphicx} % Required for inserting images
\usepackage{amsmath} % For aligning equations 
\usepackage{amssymb} % For \gtrsim
\usepackage[export]{adjustbox} 
\usepackage[T1]{fontenc}
\usepackage{bbold}
\usepackage[urlcolor=blue,colorlinks=true,citecolor=blue,linkcolor=blue,pdfstartview={FitH},bookmarks=false]{hyperref}
\usepackage{enumitem}
\usepackage{xcolor}

\usepackage[utf8]{inputenc}
\usepackage{booktabs} % For \toprule, \midrule, \bottomrule
\usepackage[percent]{overpic}

\usepackage{array}         % For column customization

\newcommand{\folderOK}{}

\begin{document}
\title{Phase structure of the one-dimensional $\mathbb{Z}_2$ lattice gauge theory\\with second nearest-neighbor interactions}

\author{Yeimer Zambrano}
%\email[e-mail: ]{aleksey.alekseev@amu.edu.pl}
%\homepage[\mbox{ORCID ID}: ]{https://orcid.org/0000-0001-6574-9299}
\affiliation{\mbox{Institute of Spintronics and Quantum Information, Faculty of Physics and Astronomy}, Adam Mickiewicz University in Pozna\'n, Uniwersytetu Pozna\'nskiego 2, PL-61614 Pozna\'n, Poland}

\author{Aleksey Alekseev}
%\email[e-mail: ]{aleksey.alekseev@amu.edu.pl}
%\homepage[\mbox{ORCID ID}: ]{https://orcid.org/0000-0001-5102-6647}
\affiliation{\mbox{Institute of Spintronics and Quantum Information, Faculty of Physics and Astronomy}, Adam Mickiewicz University in Pozna\'n, Uniwersytetu Pozna\'nskiego 2, PL-61614 Pozna\'n, Poland}
\author{Konrad J. Kapcia}
\email[corresponding author; e-mail: ]{konrad.kapcia@amu.edu.pl}
%\homepage[\mbox{ORCID ID}: ]{https://orcid.org/0000-0001-8842-1886}
%\homepage[\mbox{ORCID ID}: ]{0000-0001-8842-1886}
\affiliation{\mbox{Institute of Spintronics and Quantum Information, Faculty of Physics and Astronomy}, Adam Mickiewicz University in Pozna\'n, Uniwersytetu Pozna\'nskiego 2, PL-61614 Pozna\'n, Poland}

\author{Krzysztof Cichy}
%\email[e-mail: ]{krzysztof.cichy@amu.edu.pl}
%\homepage[\mbox{ORCID ID}: ]{https://orcid.org/0000-0002-5705-3256}
%\homepage[\mbox{ORCID ID}: ]{0000-0001-8842-1886}
\affiliation{\mbox{Institute of Physics, Faculty of Physics and Astronomy}, \mbox{Adam Mickiewicz University in Pozna\'n, Uniwersytetu Pozna\'nskiego 2, PL-61614 Pozna\'n, Poland}}

\author{Agnieszka Cichy}
\email[e-mail: ]{agnieszka.kujawa@amu.edu.pl}
%\homepage[\mbox{ORCID ID}: ]{https://orcid.org/0000-0001-5835-9807}
%\homepage[\mbox{ORCID ID}: ]{0000-0001-5835-9807}
\affiliation{\mbox{Institute of Spintronics and Quantum Information, Faculty of Physics and Astronomy}, Adam Mickiewicz University in Pozna\'n, Uniwersytetu Pozna\'nskiego 2, PL-61614 Pozna\'n, Poland}
\affiliation{Institut f\"ur Physik, Johannes Gutenberg-Universit\"at Mainz, Staudingerweg 9, D-55099 Mainz, Germany}

\date{\today}

\begin{abstract}
We investigate the ground-state phase diagram of a one-dimensional $\mathbb{Z}_2$ lattice gauge theory (LGT) model with hard-core bosons at half-filling, extending previous studies by including second nearest-neighbor (2NN) interactions. 
Using matrix product state techniques within the density matrix renormalization group, we compute charge gap, static structure factor, pair-pair correlation functions, and entanglement entropy for various interaction strengths and field parameters.
We analyze two representative neatest-neighbor interaction strengths ($V_1$) that correspond to the Luttinger liquid (LL) and Mott insulator (MI) phases in the absence of the 2NN interactions. 
We introduce the 2NN coupling $V_2$ and investigate its impact on the system. Our results reveal very rich behavior. 
As the 2NN repulsion increases, in the case of small $V_1$, we observe a direct transition from the LL phase to a charge-ordered insulator (COI) phase with four-site ordering pattern, whereas for large $V_1$, we observe a transition from the MI phase with two-site ordering pattern (previously found with only $V_1$ included), going through an intermediate LL region, and finally reaching the COI regime. 
Additionally, the inclusion of 2NN interactions enhances charge order and suppresses pair coherence, evidenced by sharp peaks in the structure factor and rapid decay in pair-pair correlators. 
Our work extends the well-studied phase structure of 1D $\mathbb{Z}_2$ LGT models and demonstrates the interplay between gauge fields, confinement, and extended interactions.
\end{abstract}

\keywords{lattice gauge theory, density matrix renormalization group, phase transitions, charge order, Mott insulator, Luttinger liquid, longer-range repulsion, charge gap, structure factor, entanglement entropy}%Use showkeys class option if keyword

\maketitle
%%%%%%%%%%%%%%%%%%%%%%%%%%%%%%%%%%%%%%%%%%%

\section{Introduction}

The idea of formulating gauge theories on a lattice was first introduced by Wegner in 1971, who studied the simplest case of the $\mathbb{Z}_{2}$ gauge group \cite{Wegner:1971app}. 
Building on this idea, Wilson later demonstrated in 1974 that lattice gauge theory (LGT) provides a non-perturbative framework for quantum chromodynamics (QCD) \cite{Wilson1974}. 
Since then, the $\mathbb{Z}_{2}$ gauge theory has become one of the most extensively studied models, with applications ranging from condensed matter systems \cite{IchinoseMatsui2014, LevinWen2005, ZoharCiracReznik2013} and quantum simulations \cite{Smith2018, Wiese2021, Marcos2013} to quantum chromodynamics and other areas of high-energy physics \cite{Hey2006, Petronzio1999, Gaz2025, He2025, Petreczky:2012rq}. 
Due to their ability to describe confinement, topological order, and phase transitions, LGTs provide a fundamental theoretical framework for systems governed by local symmetries \cite{Wilson1974, rothe2012lattice, kogut1975hamiltonian, zohar2017quantum}. 
In recent years, there has been growing interest in simplified versions of LGTs in lower dimensions, particularly the $\mathbb{Z}_2$ LGT, which captures essential features while remaining numerically tractable \cite{Kogut1979, WanTchernyshyov2013, Shaw2019ZN, Kuramashi2019}.

In condensed matter physics, $\mathbb{Z}_2$ gauge fields naturally arise in models of quantum spin liquids \cite{Savary2016}, superconductivity \cite{IchinoseMatsui2014}, and fractionalization phenomena \cite{SenthilFisher2000}. 
Moreover, recent advances in quantum simulation platforms, such as ultracold atoms and Rydberg arrays, have made possible the experimental realization of synthetic gauge fields  \cite{Dalmonte2016, zohar2017quantum}.

One-dimensional (1D) $\mathbb{Z}_2$ LGT models coupled to matter fields have attracted particular attention due to their rich phase structure and exact resolvability in certain limits. Notably, Kebri\v{c} et al. \cite{Kebric2023} mapped out the phase diagram of a 1D $\mathbb{Z}_2$ gauge theory with hard-core bosons and nearest-neighbor (1NN) density-density interactions at the half-filling, identifying two phases, a Luttinger liquid (LL) phase and a Mott insulator (MI) phase. 
Inspired by these advancements and the need to explore more scenarios, in this work, we extend the model studied in Ref.~\cite{Kebric2023} by including second nearest-neighbor (2NN) density-density interactions. 
Our primary motivation is to systematically investigate how these extended interactions modify the ground-state phase diagram and influence the correlation functions. 

To explore these effects, we employ the density matrix renormalization group (DMRG) algorithm within a matrix product state (MPS) framework \cite{white, Schollwoeck2005, schollwock11, Verstraete2008, Orus2014annphys}, we calculate the charge gap, static structure factor, and pair-pair correlators as functions of the field strength $h$ in addition to the interaction strengths $V_1$ and $V_2$, for 1NN and 2NN couplings, respectively.

Our study demonstrates that the inclusion of 2NN interactions leads to a richer phase diagram, giving rise to a novel charge-ordered insulator (COI) phase, which is characterized by a four-site charge ordering pattern that is signaled by prominent peaks in the static structure factor $S(k)$ at $k=\pi/2$ and $k=3\pi/2$. 
The inclusion of 2NN interactions in this regime suppresses quantum fluctuations, leading to the exponential decay of the pair-pair correlation function, indicating a robust charge order and minimal pair coherence. 
This COI phase does not appear in the 1NN model \cite{Kebric2023} and arises only when 2NN interaction is strong enough.
In particular, presented results reveal a phase transition from MI to COI, with an intermediate LL regime appearing between them. 
Compared to the 1NN model in Ref.~\cite{Kebric2023}, our extended model reveals how longer-range interactions shift phase boundaries, enhance charge order, and increase frustration effects linked to confinement. 
By examining two representative values of $V_1$, we show that 2NN interactions stabilize the COI phase and sharpen the transitions from metallic behavior to ordered insulating states. These findings deepen the understanding of correlated phases in lattice gauge theories (at the half-filling) and may inform future experimental efforts using tunable quantum simulators.

The remainder of this work is organized as follows.
Section \ref{hamiltonian-S} introduces the extended Hamiltonian and numerical methods, Section \ref{observables-S} defines the physical observables used for phase characterization, and Section \ref{results-S} presents and discusses our comprehensive numerical results, including the phase diagrams. 
Finally, Section \ref{conclusions-S} is devoted for conclusions and final remarks.

\section{Model and methods}\label{hamiltonian-S}

\subsection{$\mathbb{Z}_2$ lattice gauge theory Hamiltonian}
The one-dimensional $\mathbb{Z}_2$ lattice gauge theory is a simplified theoretical model widely used in field theory and condensed matter physics. In LGTs, spacetime is discretized, and the fields defined on the lattice represent the fundamental degrees of freedom of the theory: matter fields reside on the lattice sites (arranged in a line), while gauge fields live on the links connecting them. In the Hamiltonian formulation, the dynamics of the gauge fields is governed by the gauge-matter coupling and electric field terms that enforce local $\mathbb{Z}_2$ gauge invariance.
Although in higher-dimensional LGTs gauge dynamics can be expressed through plaquette terms corresponding to closed loops, in one dimension the theory lacks such spatial loops, and the dynamics is instead captured entirely by link and site interactions.

We consider a $\mathbb{Z}_2$ LGT with nearest- and second (i.e., next-) nearest-neighbor interactions between $\mathbb{Z}_2$-charged $U(1)$ matter particles, which is described by the following Hamiltonian: 
\begin{eqnarray}
\label{Hamiltonian}
H &=& -t \sum_{i=1}^{L-1} (a_i^\dagger \tau_{i,i+1}^z a_{i+1} + \text{H.c.}) - h \sum_{i=1}^{L-1} \tau_{i,i+1}^x \nonumber \\ 
&+& 
 V_1 \sum_{i=1}^{L-1} n_i n_{i+1} + 
 V_2 \sum_{i=1}^{L-2} n_i n_{i+2},
\end{eqnarray}
where $a_i^{\dagger}$ ($a_i$) is a hard-core boson creation (annihilation) operator at site $i$, $\tau^z_{ij}$ is the $z$-Pauli matrix corresponding to the $\mathbb{Z}_2$ gauge field on the link between sites $i$ and $j$, $\tau^x_{ij}$ is $x$-Pauli matrix representing the electric field on a link, and $n_i=a_i^{\dagger}a_i$ is the number operator.
The hopping parameter $t$ is taken as unity and sets the scale, $h$ is an external electric field (introducing a linear confining potential) and $V_1,V_2$ are interaction strengths between 1NN and 2NN, respectively.
As hinted in the introduction, the presence of the 2NN interaction is an extension with respect to Ref.~\cite{Kebric2023}, and we concentrate on how it affects the model properties and how it enriches the phase structure.
Finally, we mention that the gauge symmetry translates into a Gauss law \cite{ProskoPRB2017,Kebric2023}, which reads, for site~$i$,
\begin{equation}
\label{Gauss-const}
G_i=(-1)^{n_i} \tau_{i-1,i}^x\tau_{i,i+1}^x=\pm 1,
\end{equation}
and we choose to work in the physical sector $G_i=+1$ without loss of generality (and with $\tau_{0,1}=1$).

Moreover, the open boundary conditions are assumed and therefore we start and end our lattice with a link variable. Note also that in the present work all studies are performed for model (\ref{Hamiltonian}) in the half-filling case.

\subsection{Matrix product states and the density matrix renormalization group method}

In principle, the best way to investigate a discrete finite system and get access to all its eigenstates and eigenvalues is to apply the exact diagonalization (ED) of the Hamiltonian. 
The method implies that the Hamiltonian is represented in a matrix form and then diagonalized. 
In this way, one has access to all relevant physical observables. However, obviously, ED is a suitable approach only for very small systems due to an exponential growth of the Hilbert space.
Thus, to get to the thermodynamic limit, more sophisticated methods are required.
In this paper, we use the tensor network (TN) approach \cite{Verstraete2008, schollwock11, white, peps04, mera07, Orus2014annphys, Silvi2019tn, Okunishi2022, Banuls2023} and we benchmark our results against ED results for small system sizes.

Over the past decades, the TN methods underwent a significant development, which raised the numerical analysis of highly-correlated quantum many-body systems to a new level. 
Due to certain computational advantages, TNs allow one to study the properties of the systems at low temperatures, namely to efficiently obtain the ground state and low-lying excitations and perform further calculations of all relevant physical observables. 
First of all, the problem of the exponential growth of the Hilbert space, unavoidable in ED, is solved by exploiting the low entanglement of physically relevant quantum states.
TNs represent quantum states as a network of interconnected tensors, allowing for efficient computations and, by construction, do not have the sign problem that plagues Monte Carlo simulations of fermionic systems.

One of the most prominent methods in condensed matter physics has been the density matrix renormalization group (DMRG) \cite{Schollwoeck2005, white}.
Accompanied by the matrix product state (MPS) \cite{Fannes1992, Vidal2004, Verstraete2004} representation of the wave function, it became a reliable and powerful tool for solving numerous 1D quantum many-body problems. 
Here, we briefly outline the basics of the MPS method.
The MPS ansatz for a state $|\Psi\rangle$ of an $L$-site system has the form 
\begin{equation}\label{iterative}
    |\Psi\rangle = 
    \sum_{s_1=1}^d \ldots \sum_{s_L=1}^d
    \textrm{tr}\left(A_1^{s_1}\ldots A_{L}^{s_{L}}\right)
    |s_1\ldots s_{L}\rangle,
\end{equation}
where $|s_1 \ldots s_{L}\rangle$  are $d^L$ basis states of the many-body system (with $d$ being the dimension of the single-site Hilbert space) and each $A_i^{s_i}$ is a $D$-dimensional matrix, where $D$ is called the bond dimension. 
Eq. (\ref{iterative}) can be treated as a variational ansatz and the ground state can be found 
by iteratively minimizing the energy
$\langle\Psi|H|\Psi\rangle / \langle\Psi|\Psi\rangle$ with respect to each tensor $A_i^{s_i}$, until convergence is achieved.
The calculation of $\langle\Psi|\Psi\rangle$ and $\langle\Psi|\hat{O}|\Psi\rangle$ can always be done efficiently, i.e.\ at a polynomial cost in $D$ and $L$; for expectation values of operator $\hat{O}$ this is true if it is expressed as a matrix product operator (MPO) \cite{Pirvu_2010}.
Having the ground state, one can find expectation values of operators of interest \cite{white}.
To find the first excited state, one can project out the ground state and again look for the ground state of such projected Hamiltonian.
Continuing in this manner, one can find as many low-energy excited states as desired.

To apply the MPS approach, it is, in principle, enough to have the Hamiltonian as in Eq. (\ref{Hamiltonian}). 
However, it is more convenient to express the hard-core boson operators as spin ones by means of the following transformation (cf. also \cite{ProskoPRB2017,Kebric2023,Giamarchi2004,Jordan:1928wi}):
\begin{eqnarray}
a_i^{\dagger}=\sigma^+_i=\frac{\sigma^x_i+i\sigma^y_i}{2}&,&
\,\,\,\,\, 
a_i
=\sigma^-_i=\frac{\sigma^x_i-i\sigma^y_i}{2},
\nonumber\\
n_i=a_i^\dagger a_i &=& \frac{1+\sigma^z_i}{2}\equiv\pi^0_i.
\end{eqnarray}
Then, the Hamiltonian (\ref{Hamiltonian}), for a generic range of the interaction $r_{\rm max}$ (in our work, $r_{\rm max}=2$), is given by
\begin{eqnarray}
H&=&-t\sum_{i=1}^{L-1} (\sigma_i^{+} \sigma_{i+1}^{-}+\sigma_i^{-} \sigma_{i+1}^{+})-h\sum_{i=1}^{L-1}(-\sigma_1^z)\dots (-\sigma_i^z)\nonumber \\
&+& \sum_{r=1}^{r_{\rm max}} V_r \sum_{i=1}^{L-r}\pi^0_i\pi^0_{i+r}+2\lambda \sum_{i=1}^{L-1}\sum_{j=i+1}^{L}\pi^0_i\pi^0_j\nonumber \\
\label{eq:ham.MPO}
&+& \lambda(1-2N_{\rm target}) \sum_{i=1}^{L} \pi^0_i +\lambda N^2_{\rm target},
\end{eqnarray}
where we also introduced a penalty term to restrict the calculations to a fixed filling.
In our work, we consider the half-filled case, i.e., $N_{\rm target}=L/2$. The penalty term coefficient $\lambda=100$ enforces this choice in a numerically stable manner.

The 1NN+2NN model (\ref{Hamiltonian}) is, thus, equivalent to a spin system (the XXZ model (\ref{eq:ham.MPO})) in a magnetic field (associated to chemical potential) with parameters given by $h$, $V_1$ and $V_2$ (in particular, $\sigma^z_i\sigma^z_{i+r}$ are 1NN and 2NN ($r=1,2$) interactions) and additional multi-site interactions involving strings of $\sigma^z_i$ operators (up to the whole length of the system), reflecting the influence of the electric field \cite{Kebric2023,Giamarchi2004}.
The restriction to a fixed filling is realized by additional $\sigma^z_i\sigma^z_j$ interactions between all sites~$i,j$.

\section{Physical observables and phase characterization}
\label{observables-S}

\subsection{Charge gap}

The charge gap quantifies the energy cost of changing the particle number and is a fundamental indicator of whether a system exhibits metallic (gapless) or insulating (gapped) behavior. 
It has become a standard diagnostic observable in the study of interacting quantum systems, particularly in one dimension.

We are interested in even particle numbers, since unpaired particles would produce edge effects by localizing at the boundary.
Hence, we consider changing the particle number by $2$ in a fixed volume given by $L$.
The charge gap is then defined as the energy cost of adding or removing a pair of particles from the system,
\begin{equation}
 \small
    \label{Charge-Gap-EQ}
    \Delta(N, L) = \frac{E(N+2, L) + E(N-2, L) - 2E(N, L)}{2},
\end{equation}
where $E(N, L)$ is the ground-state energy of the system with $N$ particles and size $L$. This quantity vanishes in metallic phases (such as LL), while remaining finite in insulating regimes (like MI or COI). 
To obtain reliable values in the thermodynamic limit, we perform finite-size scaling of $\Delta(N, L)$ using several system sizes and extrapolate keeping $N/L$ fixed to $1/2$ (the half-filling case), obtaining $\Delta = \lim_{L \to \infty} \Delta(L/2, L)$.

\subsection{Static structure factor}
\label{Structure-Factor-Section}

The static structure factor $S(k)$ is defined as the Fourier transform of the density-density correlation function \cite{Chaikin1995, Ashcroft1976, Chanda_SF},
\begin{equation}
    \label{Structure-Factor-EQ}
    S(k) = \frac{1}{N_{\rm target}} \sum_{j,\ell} e^{ik(j - \ell)} \langle n_j n_\ell \rangle.
\end{equation}
It measures how particle densities at different sites are correlated in momentum space. Peaks in $S(k)$ indicate enhanced correlations at specific wavelengths, which correspond to the development of charge ordering with characteristic periodicity. 
It is a key observable used to detect spatial ordering and density correlations in quantum many-body systems. 
In particular, it is sensitive to the emergence of long-range charge order, making it especially useful for distinguishing between disordered metallic phases and ordered insulating phases.

In our model, the static structure factor $S(k)$, provides critical insight into the system's phase structure by mapping spatial correlations. 
In the LL phase, $S(k)$ might exhibit a broad maximum at $k=\pi$ and/or strong peaks at $k=\pi/2$ and $k=3\pi/2$~\cite{Kebric2023}.
As the interactions strengthen and the system enters insulating phases, $S(k)$ develops pronounced peaks that act as fingerprints for the specific type of charge ordering. 
The MI phase is unambiguously signaled by a prominent, sharp peak at $k=\pi$, corresponding to a two-site unit cell order. 
Conversely, the COI phase is characterized by the emergence of sharp peaks at $k=\pi/2$ and $k=3\pi/2$ with shallow minimum at $k=\pi$, indicative of a four-site ordering pattern.

By tracking the evolution of these peaks across parameter ranges (like increasing the electric field $h$ or $V_2$), one can locate phase boundaries.
Crucially, while $S(k)$ is an excellent probe of spatial order, it is not always a definitive classifier. 
The similarity in $S(k)$ between the LL and COI phases suggests that observables like the charge gap ($\Delta (N, L)$) are required to robustly distinguish between metallic (gapless) and insulating (gapped) states.

\subsection{Pair-pair correlation function}

Pair-pair correlation functions are used to probe the presence and nature of coherence between bound pairs of particles in interacting many-body systems. 
In the $\mathbb{Z}_2$ LGT, where hard-core bosons are coupled to $\mathbb{Z}_2$ gauge fields, a gauge-invariant pairing operator must include appropriate string operators composed of $\hat{\tau}^z$ gauge links. 
We define the gauge-invariant one-length pair-pair correlation function between sites $i$ and $j$ as \cite{Borla2020, Kebric2023}
\begin{equation}
\label{pair_corr}
\langle b_i^{\dagger} b_j \rangle = \left\langle
a^{\dagger}_{i} \tau_{i, i+1}^{z} a^{\dagger}_{i+1} 
a_{j} \tau_{j, j+1}^{z} a_{j+1}
\right\rangle.
\end{equation}
Here, the correlation depends only on the separation $r = |i-j|$ between sites $i$ and $j$; in our calculations, for a given $r$, we average the results over all pairs $(i,j)$ with the same separation to improve statistical accuracy and reduce boundary effects.
Gauge-invariant pair correlators like those have been used in previous studies of LGTs coupled to matter, where confinement or string-breaking effects alter propagation properties of pairs \cite{zohar2017quantum, Barbiero2019}.

The LL phase has quasi-long-range order, i.e., algebraic decay, while insulators (the MI and COI phases) have exponential decay of correlations related to the presence of the charge gap.
In the context of our model, one can expect a transition from algebraic to exponential decay at boundaries between the LL and insulator phase lines, but like in the case of the charge gap, the character of the decay does not determine whether it is the MI or the COI phase.

\subsection{Entanglement entropy and central charge}\label{sec:entandcetral}

After obtaining the ground state via DMRG simulations, we proceed to analyze the von Neumann entanglement entropy ($S_{vN}$). 
This observable quantifies the quantum correlations between two bipartite halves of the chain. For each bond $b$ along the chain, the system is effectively partitioned into a left subsystem (sites $1, \ldots, b$) and a right subsystem (sites $b+1, \ldots, L$). 
The $S_{vN}$ at bond $b$ is then calculated from the singular values obtained by performing a singular value decomposition (SVD) on the corresponding bond tensor of the MPS representation  \cite{Schollwoeck2005, NielsenChuang2000}. 
The probabilities $p_i$ associated with each Schmidt state are obtained by squaring the singular values $\lambda_i$ \cite{PerezQIC2007,schollwock11,BanulsPRD2019}: $p_i = |\lambda_i|^2$.
Then the von Neumann entanglement entropy $S_{vN} {(b)}$ for the bipartition at bond $b$ is calculated by using the formula \cite{NielsenChuang2000}:
    \begin{equation}
    \label{SvN}
        S_{vN}{(b)} = - \sum_i p_i \log(p_i).
    \end{equation}

\begin{figure}[b]
    \centering
    \includegraphics[width=1.0\linewidth]{\folderOK 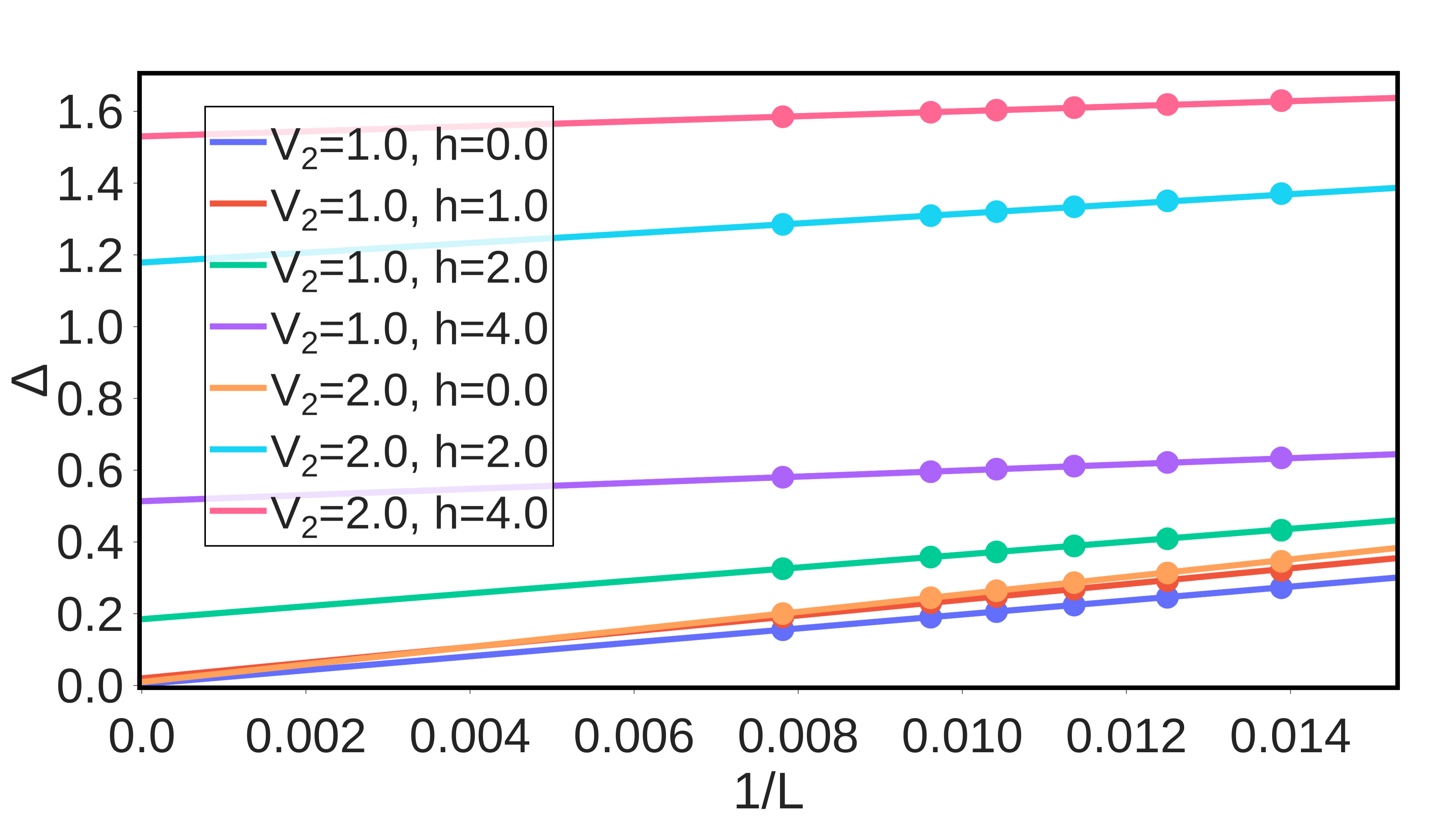}
    \caption{Finite-size scaling extrapolation of the charge gap $\Delta(L/2, L)$ for  $V_1 = 1.0$ (and different values of $h$ and $V_2$ as labeled). 
    The linear fit is performed as a function of $1/L$, and the intercept at $1/L \to 0$ yields $\Delta$.}
    \label{Scaling-CCG}
\end{figure}

In one-dimensional quantum many-body systems at criticality, the entanglement entropy provides a direct connection to the underlying conformal field theory (CFT) given by the Calabrese-Cardy equation \cite{CalabreseCardy2004}:
\begin{equation} 
\label{central-charge} 
S_{vN}(b) = \frac{c}{6} \log \left[ \left( \frac{2L}{\pi}\right) \sin \left( \frac{\pi b}{L} \right) \right] + S_{0},
\end{equation}
where $c$ is the central charge of the CFT and $S_{0}$ is a non-universal constant. 
The central charge is a fundamental parameter in conformal field theory that characterizes the number of effective degrees of freedom at criticality~\cite{DiFrancesco1997}.

\begin{figure}[t]
    \centering
    \includegraphics[width=1.0\linewidth,center]{\folderOK 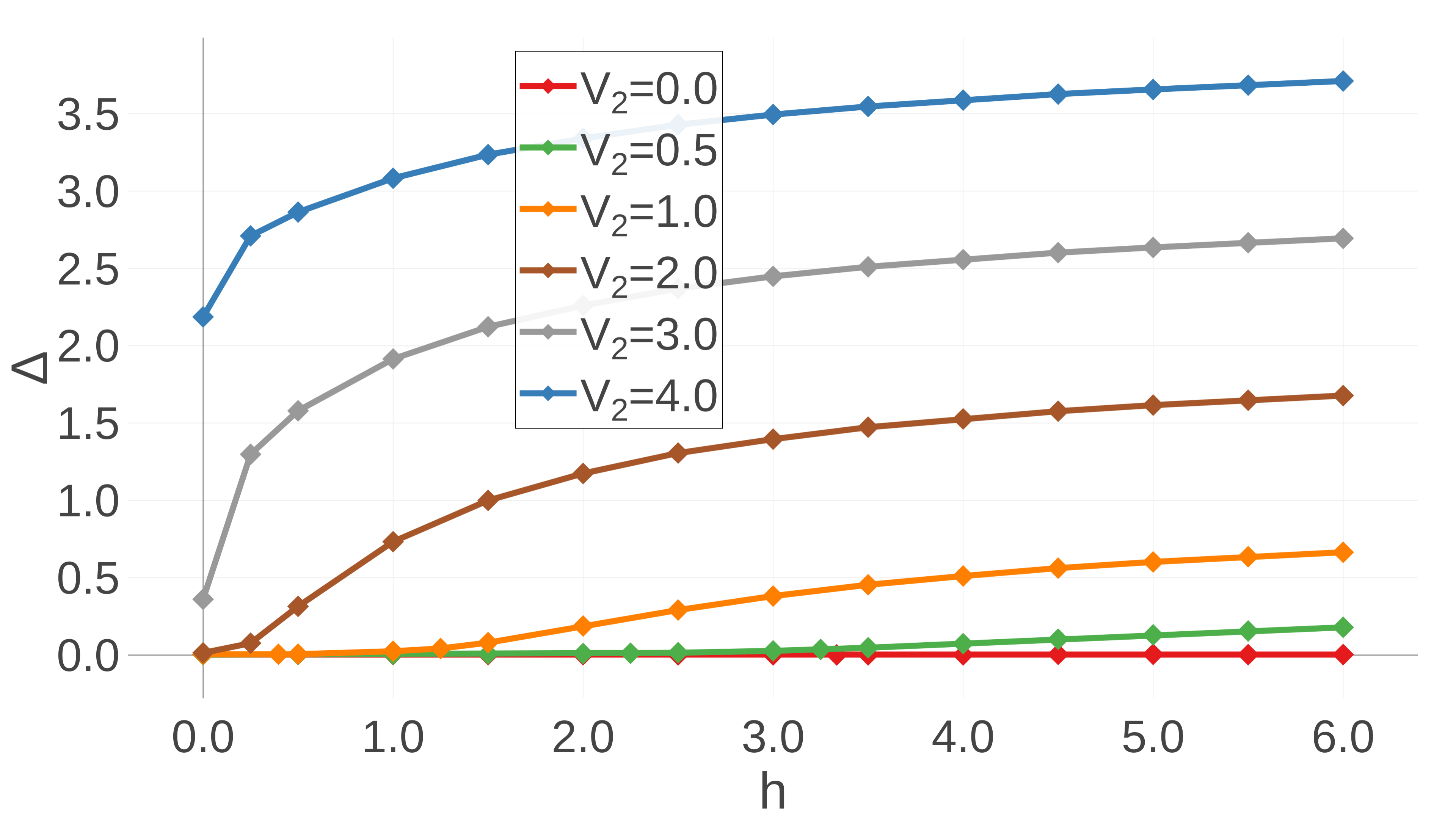}
    \includegraphics[width=1.0\linewidth,center]{\folderOK 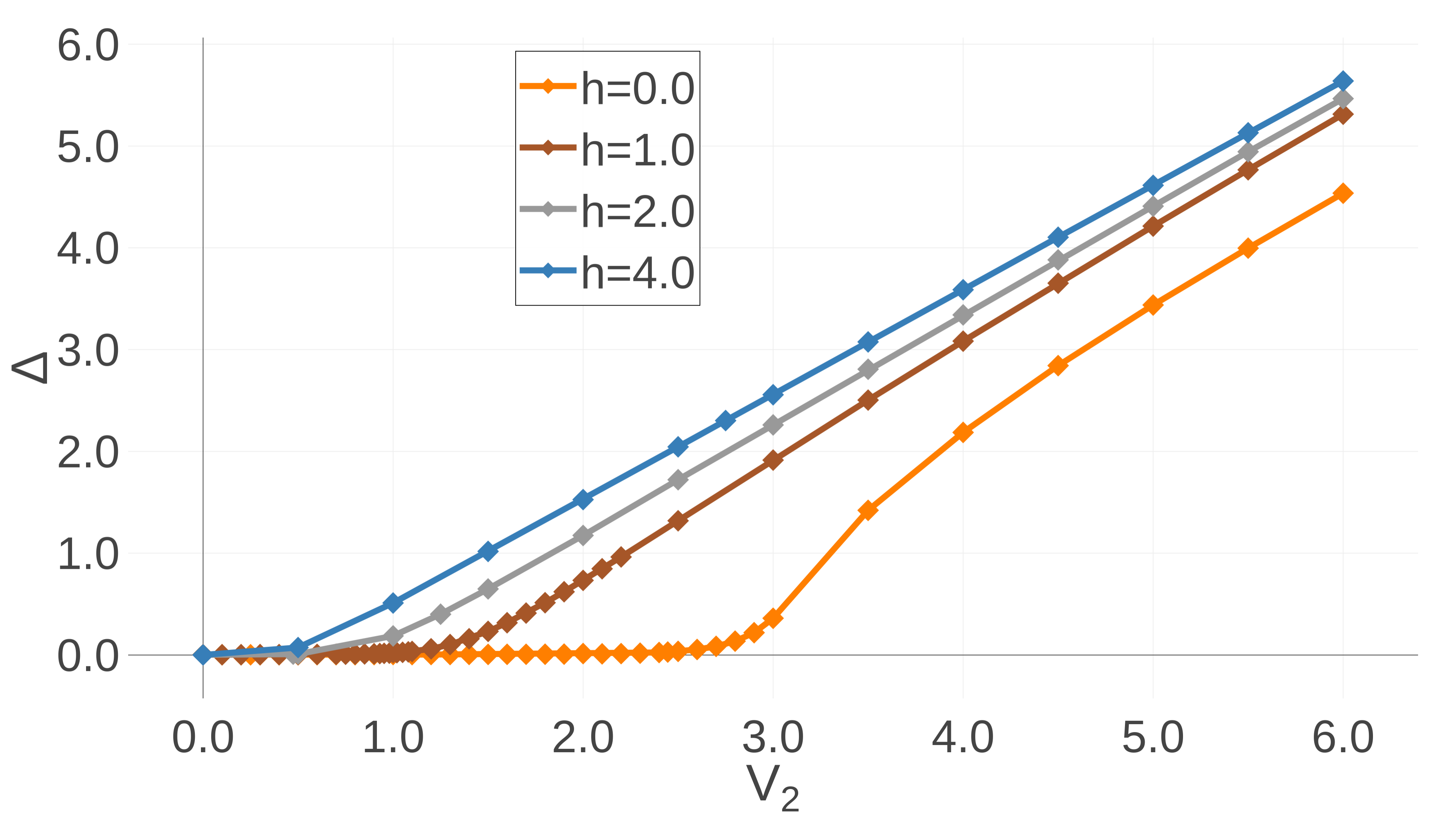}
    \caption{The critical charge gap $\Delta$ for $V_1=1.0$ presented as a function of $h$ (upper panel, for fixed $V_2$ as labeled) and of $V_2$ (lower panel, for fixed $h$ as labeled). 
    Linear extrapolation of $\Delta$ to zero identifies critical points, which define the LL-COI phase transition boundary (cf. also Sec.~\ref{Phase-Diagram-S}).}
    \label{delta-critical-v1=1}
\end{figure}

\begin{figure}[t]
    \centering    
    \includegraphics[width=1.0\linewidth,center]{\folderOK 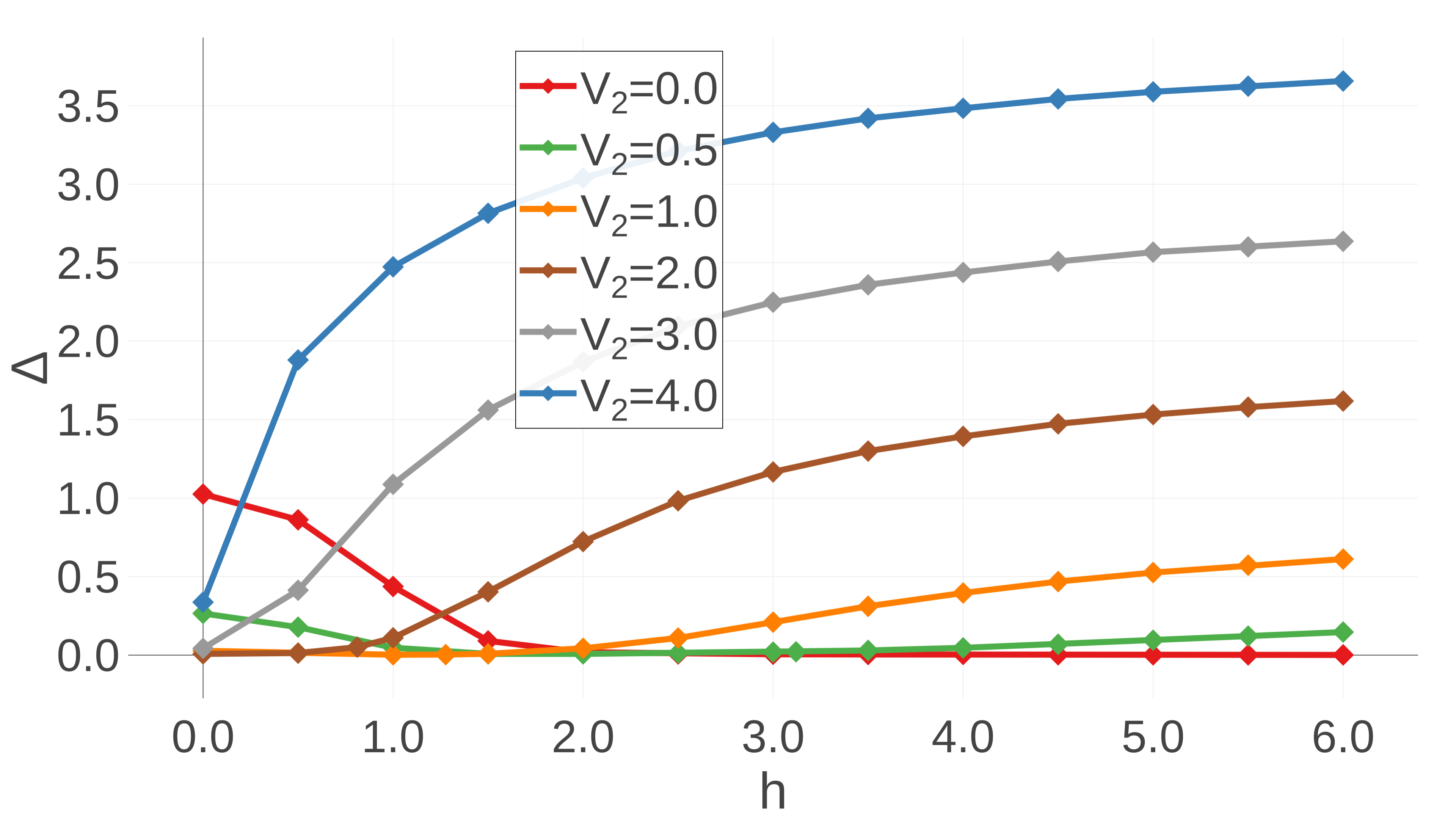}
    \includegraphics[width=1.0\linewidth,center]{\folderOK 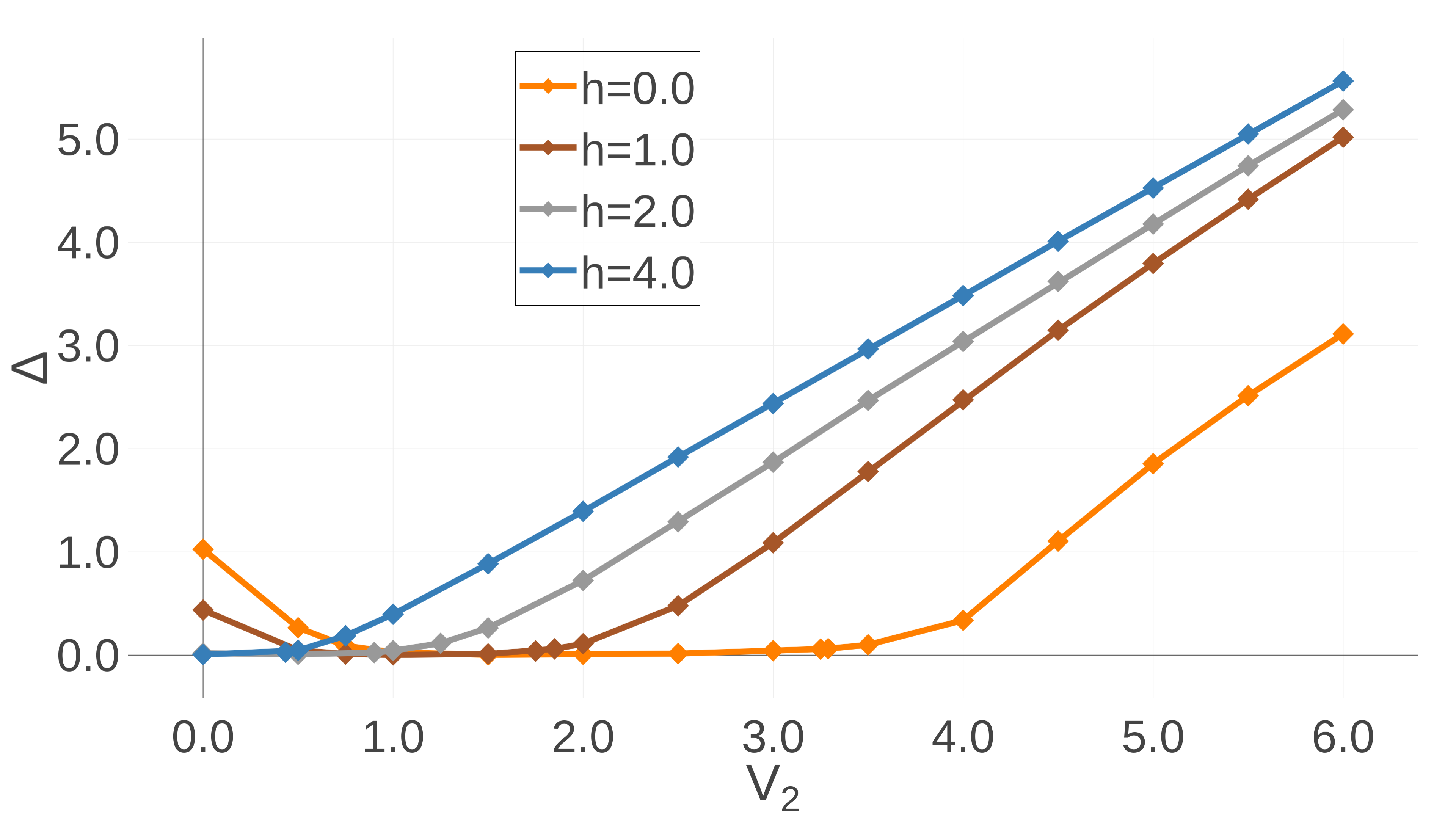}
    \caption{The critical charge gap $\Delta$ for $V_1=4.0$ as a function of $h$ (upper panel, for fixed $V_2$ as labeled) and of $V_2$ (lower panel, for fixed $h$ as labeled). 
    The points, from which phase transitions occur are identified by linearly extrapolating $\Delta$ to zero, it constitutes the basis for constructing the MI-LL and the LL-COI phase boundaries (cf. also Sec.~\ref{Phase-Diagram-S}).}
    \label{delta-critical-v1=4}
\end{figure}

For a finite system size $L$, one cannot strictly locate phase transition points and expect the Calabrese-Cardy relation to hold quantitatively, as this relation is strictly valid only at criticality and in the thermodynamic limit ($L \rightarrow \infty$). 
Instead, we utilize this analysis to seek a qualitative confirmation of the phase boundaries.

\section{Results and discussions}
\label{results-S}

\subsection{Critical charge gap $\Delta$}
\label{critical-charge-gap-S}

As mentioned previously, we extend the 1NN model (Ref.~\cite{Kebric2023}) by incorporating 2NN interactions. 
We focus on two representative values: $V_1/t = 1$ (corresponding to a LL phase in the 1NN model) and $V_1/t = 4$ (corresponding to a MI phase in the 1NN model at low $h$). 
For each $V_1$, we calculate the charge gap $\Delta(L/2, L)$ across six system sizes $L = [72, \ 80, \ 88,\ 96, \ 104, \ 128]$ and apply finite-size scaling to determine the critical charge gap $\Delta$. 
To ensure numerical reliability, we systematically increase the MPS bond dimension and verify convergence of $\Delta(L/2, L)$. 
Beyond a maximum bond dimension of approximately $500$, the results show no significant change, indicating that the simulations are well converged within this threshold.

Our procedure is as follows. 
For the specified $V_1$ values and for each system size $L$, we vary $V_2$ and $h$, computing $\Delta(L/2, L)$ according to Eq.~(\ref{Charge-Gap-EQ}). 
The thermodynamic limit of the charge gap is then obtained by plotting $\Delta(L/2, L)$ as a function of $1/L$ and performing a linear extrapolation. The extrapolated value at $1/L \to 0$ (i.e., $L \to \infty$) yields critical charge gap $\Delta$. 
This methodology is illustrated in Fig.~\ref{Scaling-CCG}.

Once $\Delta$ is obtained, we can characterize emergent phases across the $(V_2, h)$ plane. The phase transitions, indicated by the closing of the gap, are identified by analyzing $\Delta$ as a function of $V_2$ and $h$ (see Figs. \ref{delta-critical-v1=1} and \ref{delta-critical-v1=4} and Sec. \ref{Phase-Diagram-S}). 
We call an insulating state appearing for large $h$ and $V_2$ as the COI phase. In the following sections, we show that this phase has indeed different properties that insulating state occurring for small $h$ and $V_2$ (and large enough $V_1$), which is the MI phase found for $V_2=0$ previously in Ref.~\cite{Kebric2023}.

\begin{figure}[b]
    \centering    
    \includegraphics[width=1.0\linewidth,center]{\folderOK 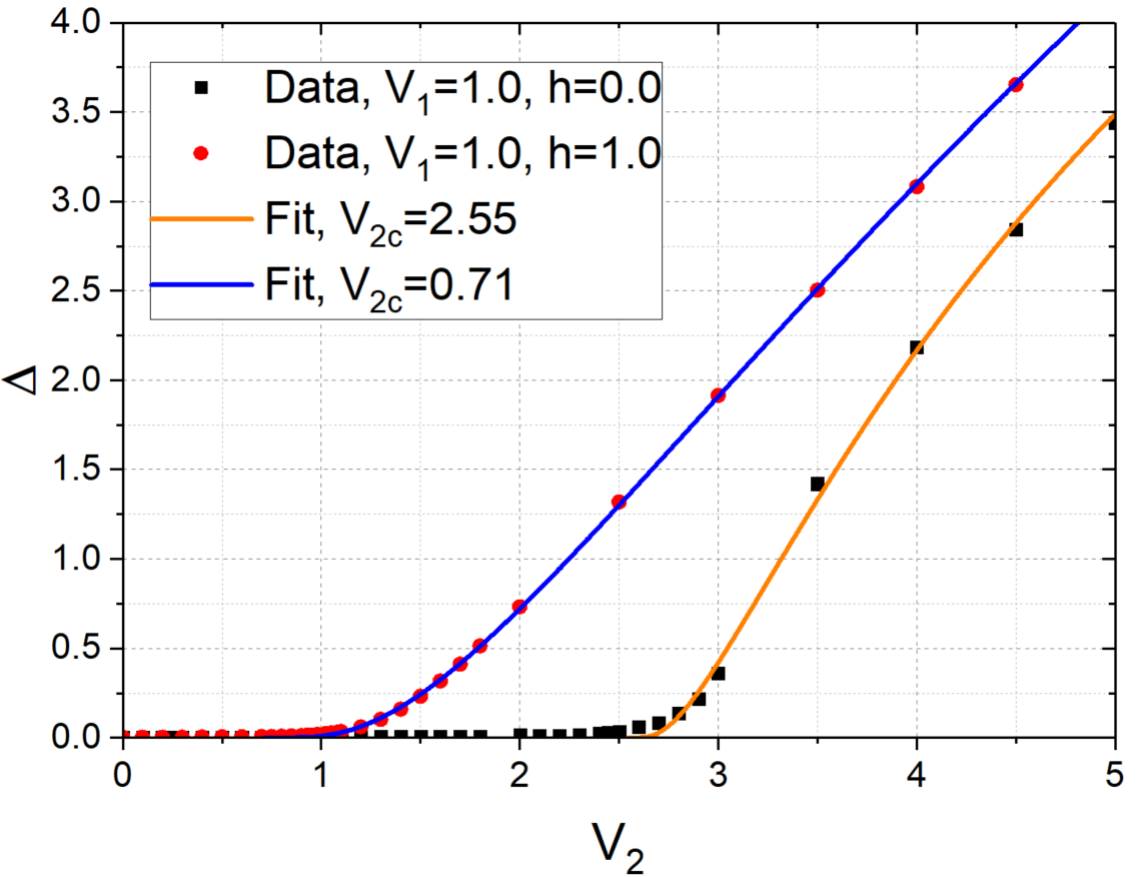}
    \caption{The critical charge gap $\Delta$ opening in the COI phase as a function of $V_2$ for $h=0.0$ (black squares) and $h=1.0$ (red circles); both for $V_1=1.0$.
    For both cases the Berezinskii-Kosterlitz-Thouless-like fit (\ref{eq:fit}) is presented (orange and blue lines, respectively).
    The values of $V_{2c}$, at which $\Delta\rightarrow 0$ is given in the figure label.}
    \label{BKTopening}
\end{figure}

However, it is worth mentioning that we performed a systematic finite-size scaling analysis of the charge gap $\Delta$ to clarify the nature of the low-$V_2$ regime in the presence of Berezinskii-Kosterlitz-Thouless-like (BKT-like) behavior. 
While extrapolations alone cannot unambiguously distinguish a truly gapless phase from an exponentially small gap, the $V_2$-dependence of $\Delta$ is consistently captured only by scaling forms of the type \cite{Kebric2023,Giamarchi2004}:
\begin{equation}\label{eq:fit}
\Delta = A \exp\left(-\frac{B}{\sqrt{V_2-V_{2c}}}\right), 
\end{equation}
with a finite critical value $V_{2c}\neq 0$ ($A$ and $B$ also treated as a fitting parameters). 
This indicates a continuous closing of the gap at the LL-COI transition and supports the existence of a finite $V_{2c}$ for both $h=0$ and $h\neq 0$.
The fit of function (\ref{eq:fit}) to the data of $\Delta$ as a function of $V_2$ obtained for $h=0.0$ and $h=1.0$ (both for $V_1=1.0$) is presented in Fig.~\ref{BKTopening}.

In particular, we carried out a systematic finite-size scaling using several fitting schemes (linear, quadratic, and cubic in $1/L$) to reliably extrapolate $\Delta(L/2,L)$ to the thermodynamic limit. 
As mentioned above, such extrapolations alone do not allow for an unambiguous distinction between a strictly gapless phase and a regime with an exponentially small but finite gap.
Instead, the dependence of the charge gap on the next-nearest-neighbor interaction $V_2$ provides the crucial insight. 
Specifically, we find that the data are consistently described by the BKT-like scaling form (\ref{eq:fit}) only when a finite critical value $V_{2c}\neq 0$ is assumed. Conversely, the fits imposing $V_{2c}=0$ fail to capture the observed behavior. 
This conclusion holds for both $h=0$ and $h\neq 0$, indicating that the transition between the Luttinger liquid and the charge-ordered insulating phase occurs via a continuous closing of the gap at a finite $V_{2c}$ (cf. Fig. \ref{BKTopening}). 
These results are further supported by the behavior of additional observables discussed in the manuscript, which consistently locate the LL-COI phase boundary at nonzero $V_{2c}$.

 \subsection{Static structure factor analysis}

As the first part of this analysis, we are interested in observing changes in $S(k)$ across all occurring phases (LL, MI, and COI) for different system sizes $L$. 
As detailed in Sec.~\ref{Structure-Factor-Section}, $S(k)$ serves as a diagnostic tool for identifying characteristic ordering patterns: a peak at $k=\pi$ signals the MI order, while the emergence of peaks at $k=\pi/2$ and $3\pi/2$ indicates COI-type charge ordering.

\begin{figure}[b]
    \centering
    \includegraphics[width=1.0\linewidth,center]{\folderOK 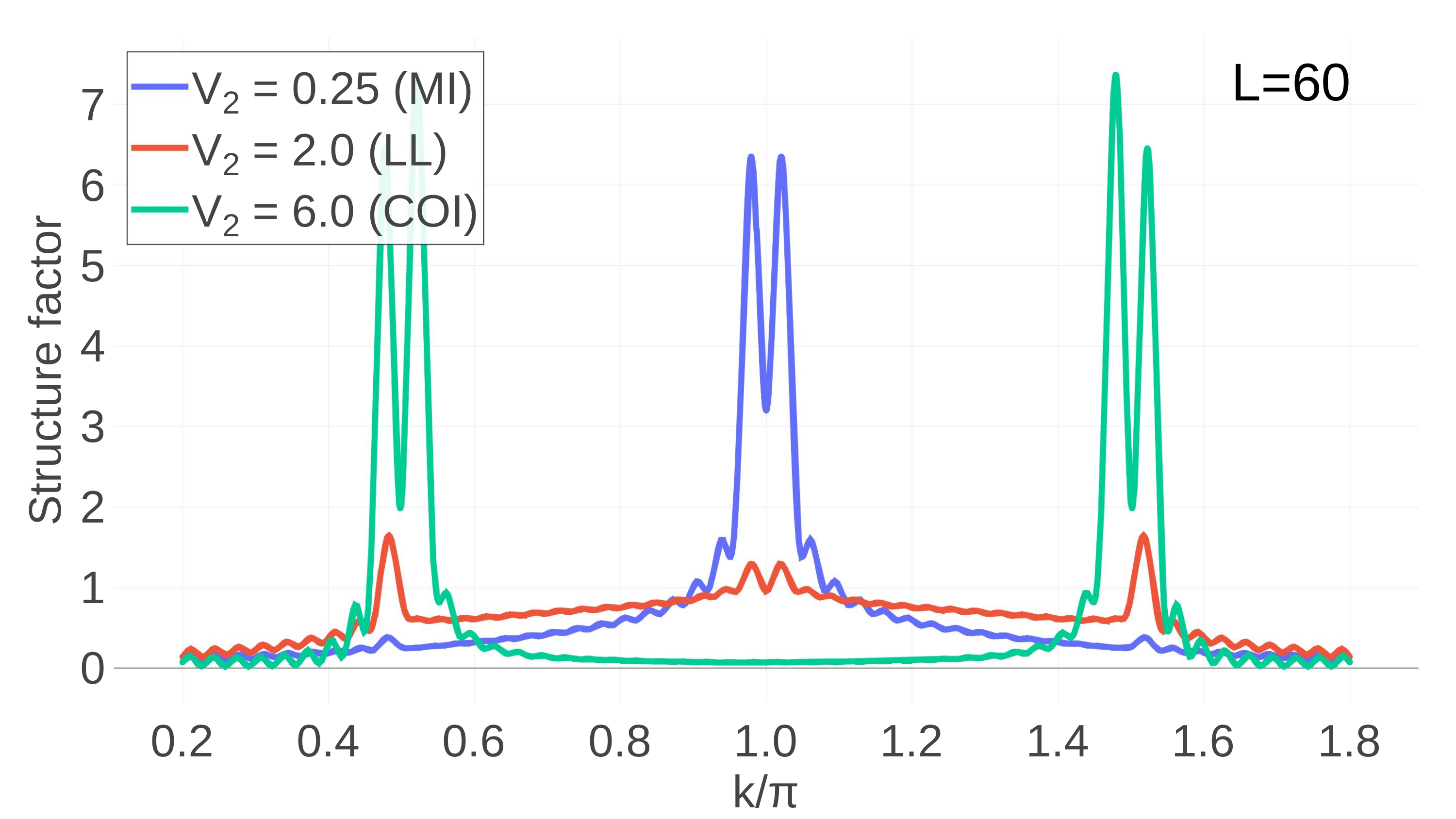}
    \includegraphics[width=1.0\linewidth,center]{\folderOK 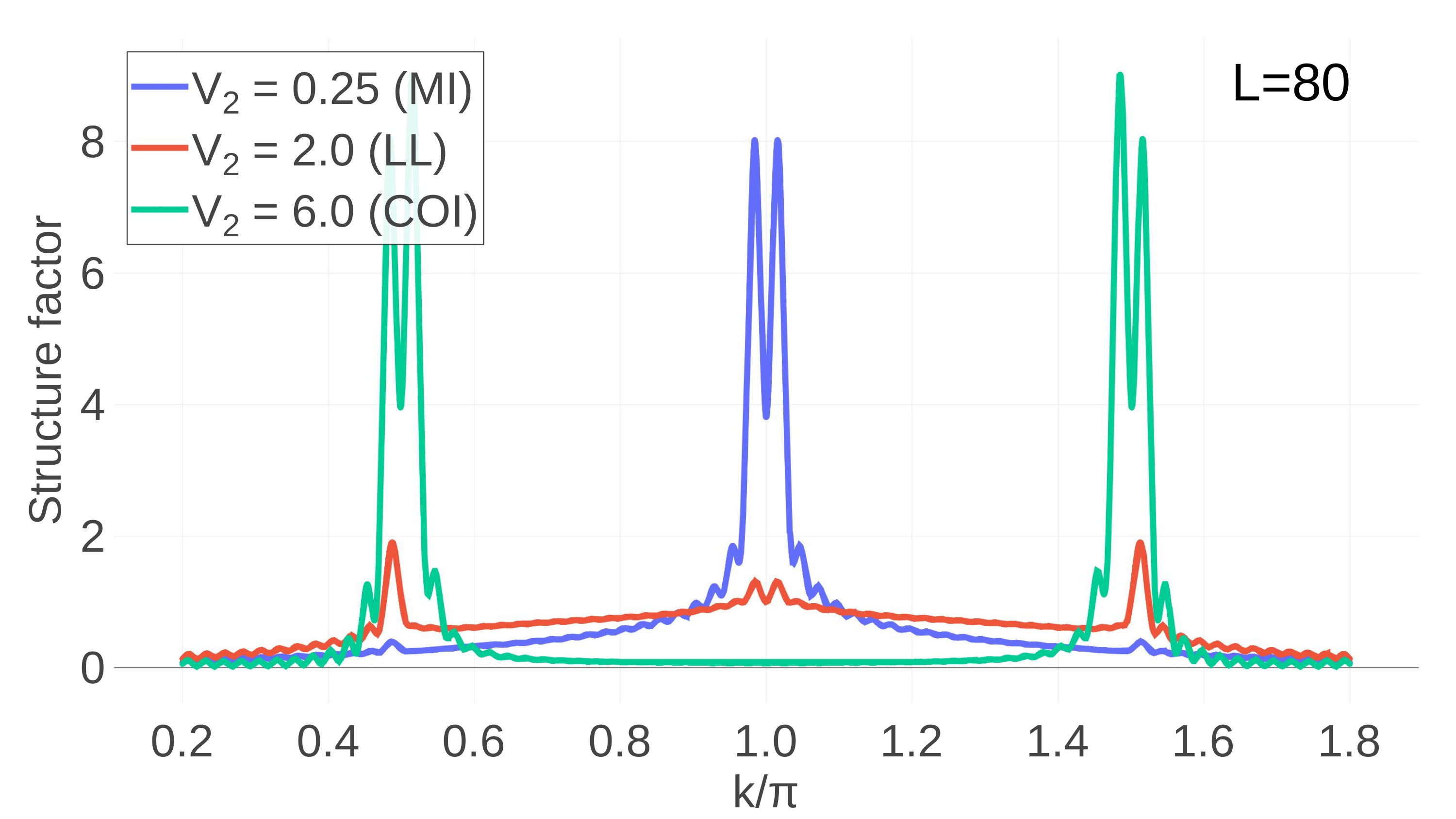}
    \includegraphics[width=1.0\linewidth,center]{\folderOK 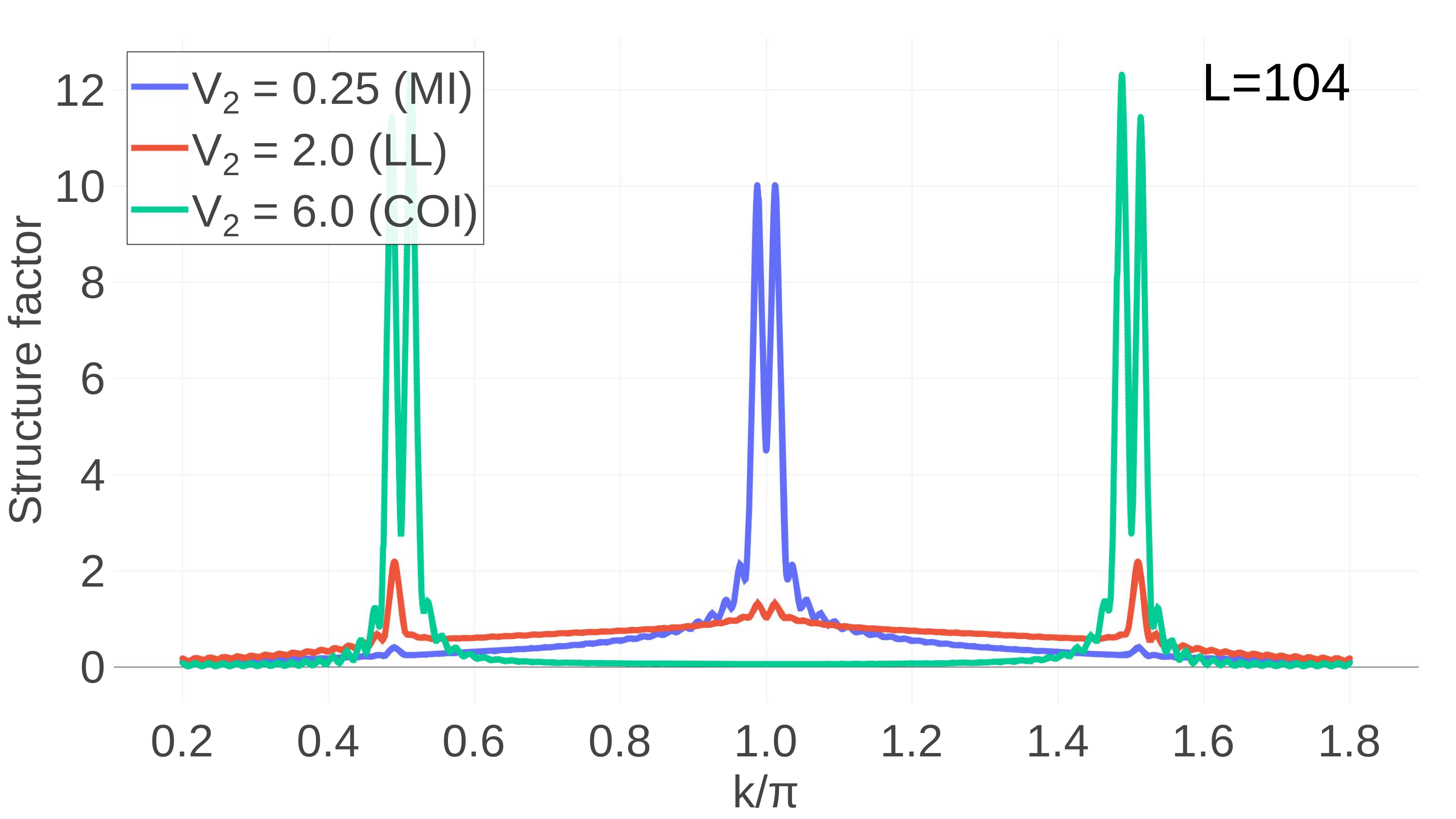}
    \caption{Static structure factor $S(k)$ for $V_{1}=4.0$ and $h=0.5$ for different system sizes: $L=60$ (upper panel), $L=80$ (middle panel), and $L=104$ (lower panel). 
     The peak at $k=\pi$ indicates the MI phase, while the peaks at $k=\pi/2$ and $3\pi/2$ signal the LL and the COI phases. 
    The comparison illustrates the finite-size convergence of $S(k)$ toward the thermodynamic limit.}
  \label{SFL60-80-104}
\end{figure}

In the $(V_2, \ h)$ plane for $V_1 = 4.0$, we take a cut along $h=0.5$ (this particular selection allows us to have all the phases to characterize, cf. Sec. \ref{Phase-Diagram-S}). 
Next, for $V_2$ = [0.25, 2.0, 6.0], we calculate $S(k)$ according to Eq.~(\ref{Structure-Factor-EQ}). 
Based on this characterization, the point $(V_2, \ h)$ = (0.25, 0.5) corresponds to the MI phase, (2.0, 0.5) to the LL phases, and  (6.0, 0.5) to the COI phases. 
The results of this part are presented in Fig.~\ref{SFL60-80-104}. 

For the MI phase (blue line in Fig.~\ref{SFL60-80-104}), a clear finite-size trend is observed as the system size $L$ increases. 
At smaller $L$, the maximum of $S(k)$ appears around $k = \pi$ as a double-peak structure. With increasing $L$, these two peaks move closer together and gradually merge into a single, sharper maximum at $k = \pi$. 
Thus, in the thermodynamic limit, $S(k)$ for the MI phase is expected to exhibit a pronounced maximum at $k = \pi$, with values approaching zero elsewhere for ${k \in (0, 2\pi)} \backslash \{ \pi \} $.

\begin{figure}[b]
  \centering
    \includegraphics[width=1.0\linewidth,center]{\folderOK 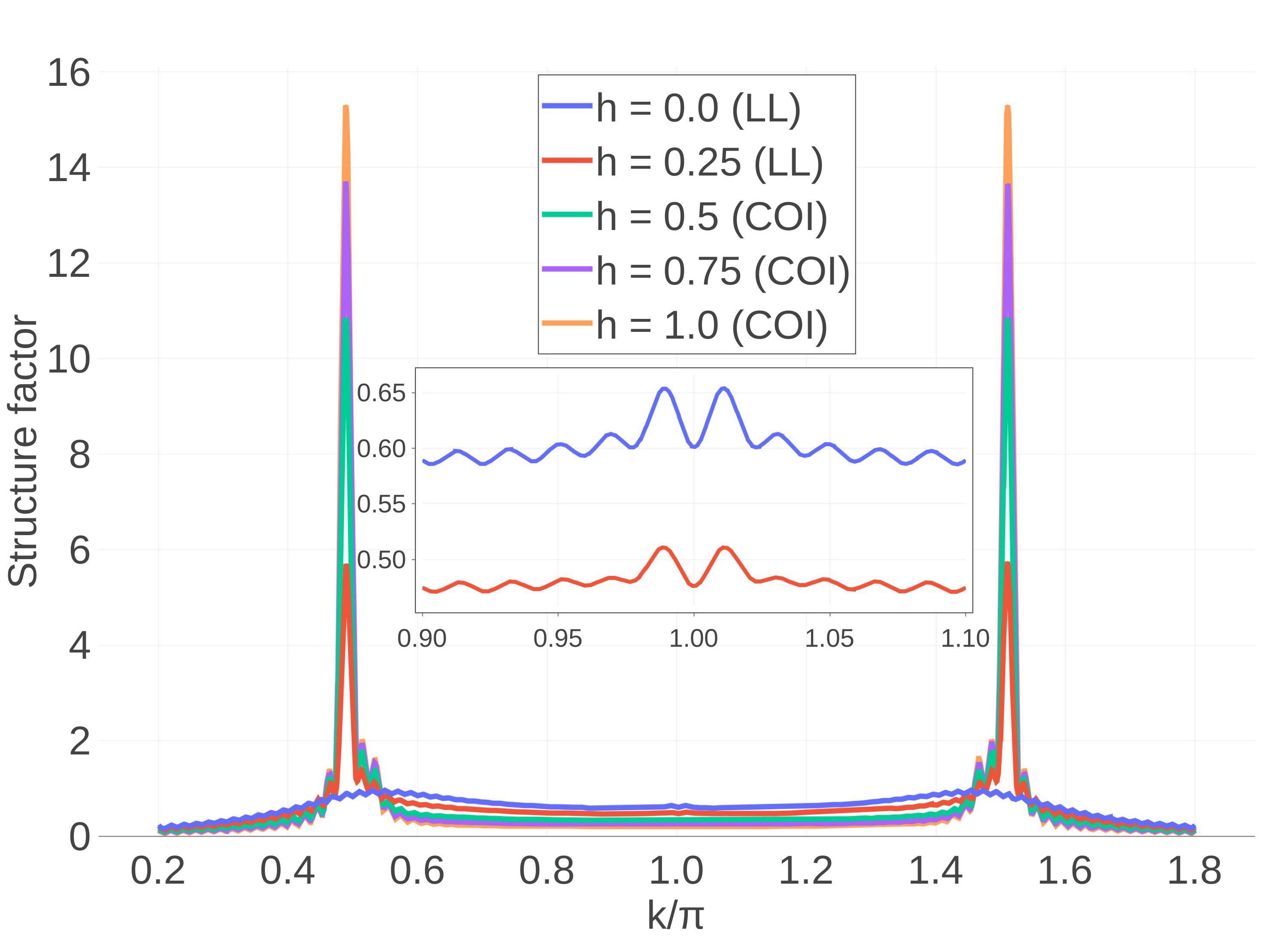}
    \includegraphics[width=1.0\linewidth,center]{\folderOK 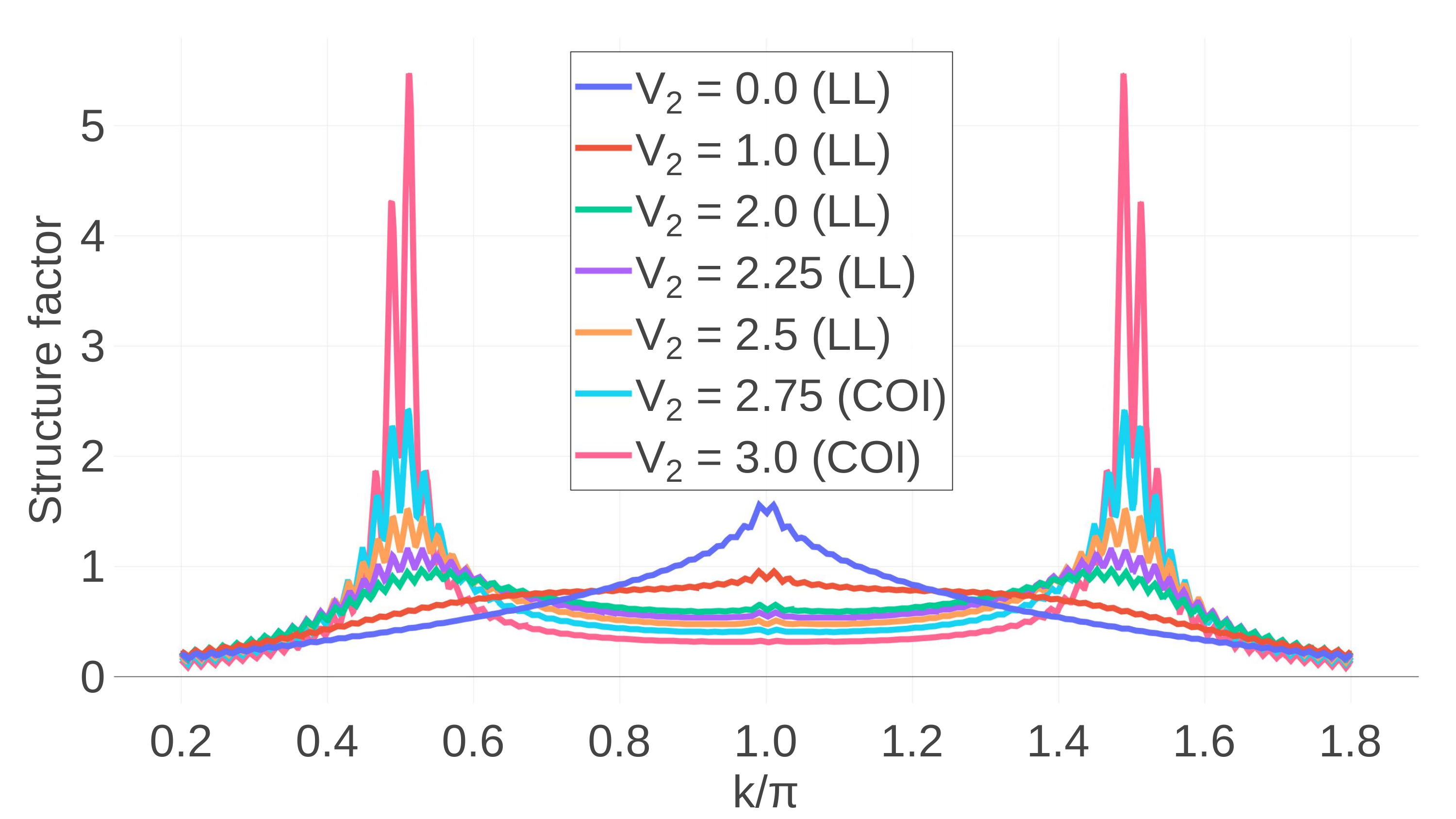}
  \caption{Structure factor $S(k)$ across the LL to the COI phase transition for $V_1 = 1.0$ and $L=104$. 
  The upper panel shows the transition driven by increasing the magnetic field $h$ (as labeled, at fixed $V_2=2.0$). 
  The lower panel shows the transition driven by increasing the long-range interaction $V_2$ (as labeled, at fixed $h=0.0$). 
  In both cases, the $S(k)$ features transition from the LL behavior to the two pronounced peaks characteristic of the COI phase.}
  \label{fig:SFL_COI_V1_1_Combined}
\end{figure}

\begin{figure}[b]
  \centering
    \includegraphics[width=1.0\linewidth,center]{\folderOK 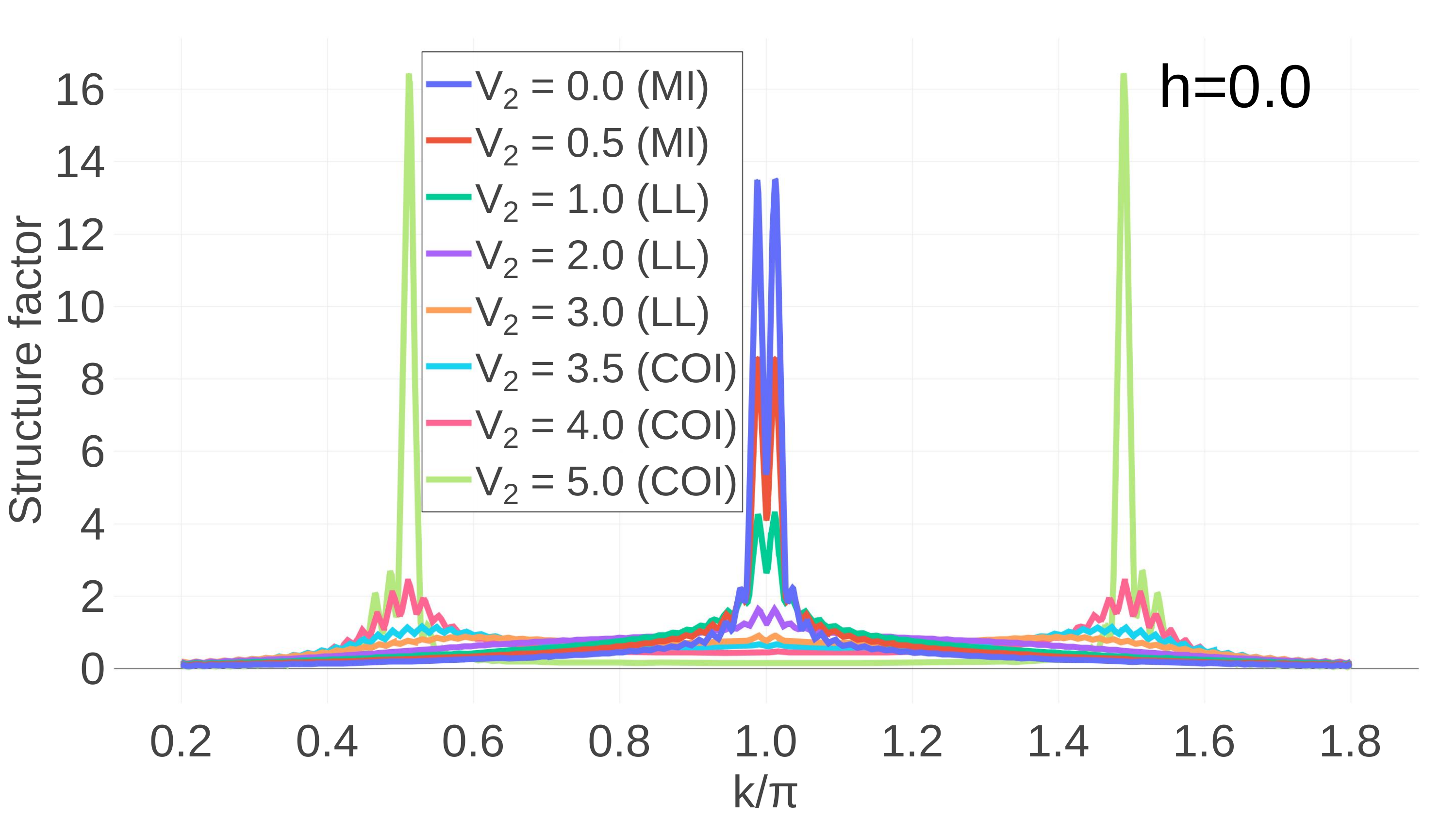}
    \includegraphics[width=1.0\linewidth,center]{\folderOK 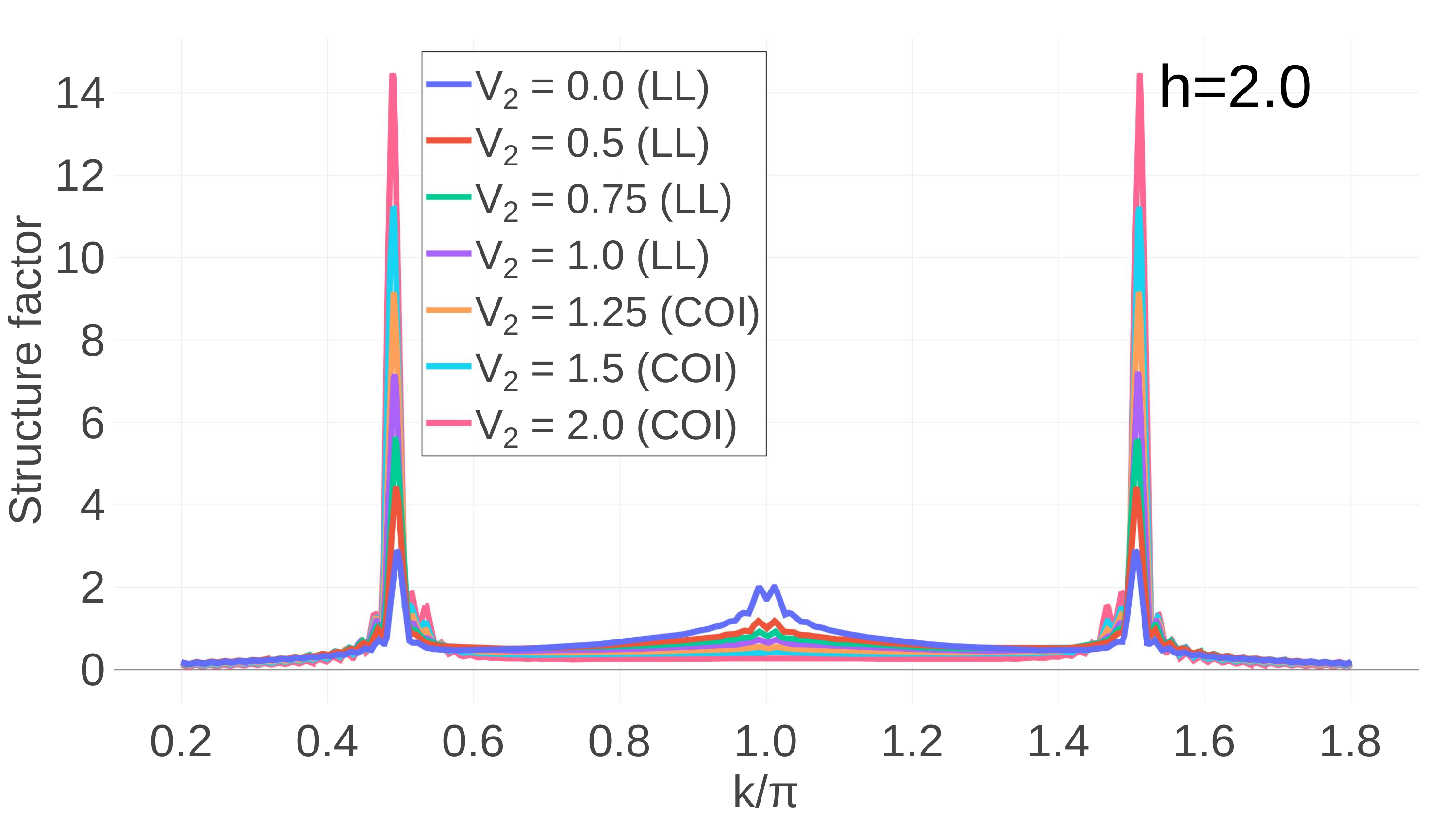}
    \includegraphics[width=1.0\linewidth,center]{\folderOK 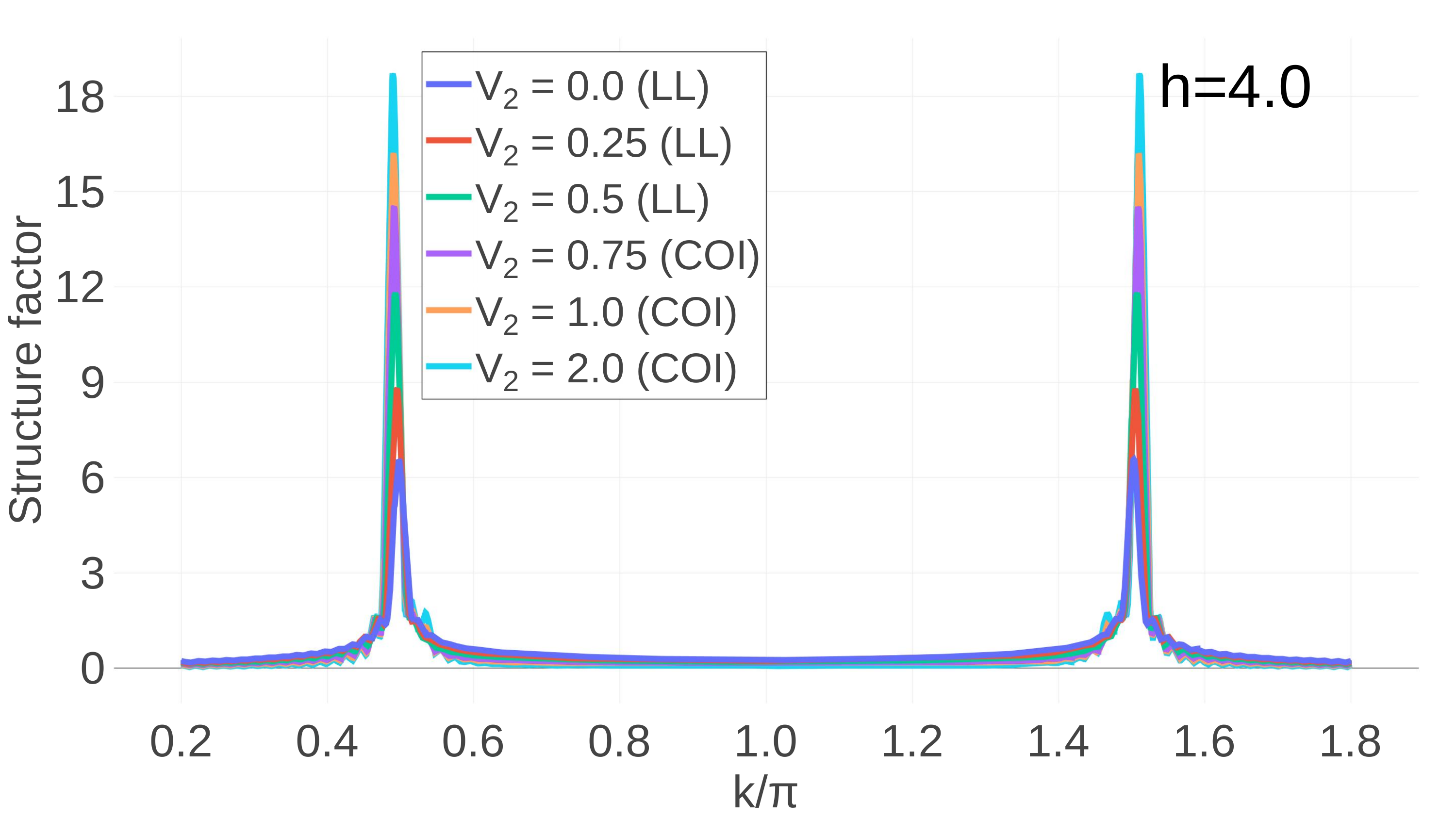}
  \caption{Evolution of the static structure factor $S(k)$ profiles for $V_1 = 4.0$ and $L=104$ as the long-range interaction $V_2$ is varied, shown for three different magnetic field cuts: $h = 0.0$ (top panel),  $h = 2.0$ (middle panel), and  $h = 4.0$ (bottom panel). Note how the distinction between the LL and the COI profiles diminishes as $h$ increases.}
  \label{fig:SF-V1-4-h-Combined} 
\end{figure}

Turning to the LL (red line) and COI (green line) phases in Fig.~\ref{SFL60-80-104}, both exhibit two primary maxima around $k = \pi/2$ and $k = 3\pi/2$. 
However, a key difference emerges: in the COI phase, $S(k)$ approaches zero for other $k$ values (shallow minimum at $k=\pi$), whereas in the LL phase, a local broad maximum can persist at $k = \pi$. 
This distinction can be attributed to the nature of particle pairing in these phases. 
Both COI and LL involve the formation of pairs. 
However, in the LL phase, these pairs are in continuous motion (delocalized), leading to a broader distribution of $S(k)$ and less pronounced maxima due to spatial averaging. 
From the other side, in the COI phase, the pairs are more localized, resulting in significantly sharper and larger maxima at $k = \pi/2$ and $k = 3\pi/2$.

While $S(k)$ effectively distinguishes the MI phase from both the LL and COI phases, it proves less definitive in differentiating between the LL and COI phases. 
The similarities in their $S(k)$ profiles suggest that the static structure factor alone is not a fully reliable criterion for robustly distinguishing between LL and COI. 
In such cases, the charge gap provides a more definitive characteristic for their differentiation.
Thus,  $\Delta$ gives the distinction between the LL phase vs. insulating phases (i.e., the MI and COI phases) and then the structure factor allows to disentangle the MI phase vs. the COI phase.

Going one step further in this analysis, we now investigate the evolution of $S(k)$ as the system transitions from the LL phase to the COI phase, setting $V_1 = 1.0$ (and for $L=104$). 
This transition is explored by varying one parameter while keeping the other fixed, which allows us to analyze two distinct mechanisms for driving the transition, as shown in Fig.~\ref{fig:SFL_COI_V1_1_Combined}.

The upper panel of Fig.~\ref{fig:SFL_COI_V1_1_Combined} illustrates the transition when $V_2$ is fixed and the magnetic field $h$ is tuned. Starting in the LL phase (small $h$), $S(k)$ exhibits a characteristic plateau (with some small oscillations). 
As $h$ increases, two small peaks begin to emerge at $k=\pi/2$ and $k=3\pi/2$. These peaks rapidly sharpen and stabilize into the pronounced, highly localized structure that defines the COI phase at larger values of $h$.

Conversely, lower panel of Fig.~\ref{fig:SFL_COI_V1_1_Combined} contrasts this behavior by showing the $S(k)$ profile as a function of the long-range interaction $V_2$ (with $h$ fixed). As $V_2$ increases, the structure factor evolves from a single broad maximum at $k=\pi$ (characteristic of the LL phase) into a minimum. 
This suppression of the $k=\pi$ peak occurs simultaneously with the emergence of the two $k=\pi/2$ and $k=3\pi/2$ peaks, demonstrating a fundamental shift in the dominant wavelength of the charge correlations to the COI ground state.

\begin{figure}[t]
    \centering
    \includegraphics[width=1.05\linewidth,center]{\folderOK 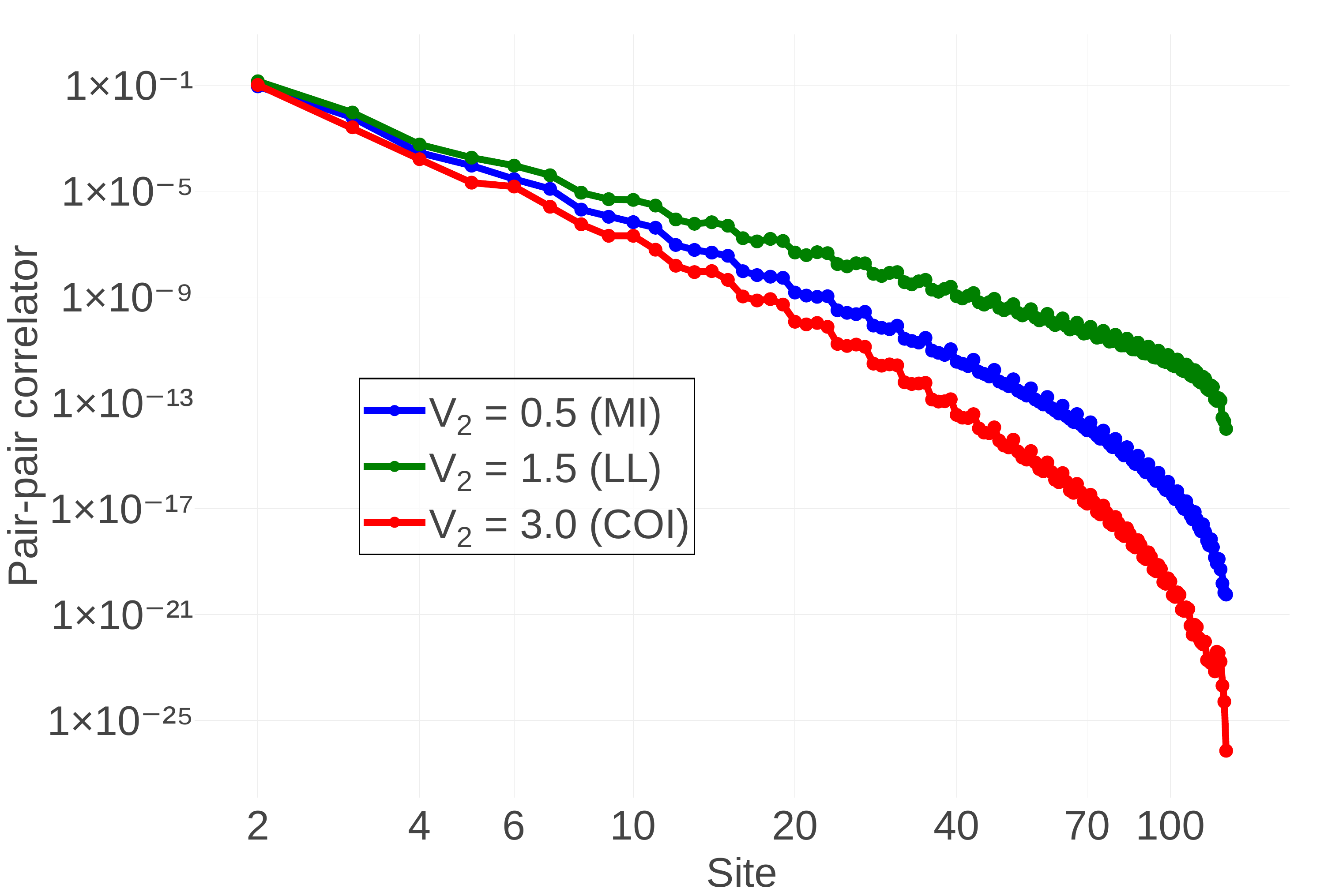}
    \caption{One-length dimer pair--pair correlators $\langle b_0^{\dagger} b_j \rangle$ for MI ($V_2 = 0.5$), LL ($V_2 = 1.5$), and COI ($V_2 = 3.0$) phases (as labeled) at $V_1 = 4.0$, $h = 0.5$, and system size $L = 128$. 
    The COI phase shows the fastest decay, consistent with strong localization and short-range order.}
    \label{PPC-MI-FL-COI-L128}
\end{figure}

Finally, for the $V_1 = 4.0$ case, analyzing the evolution of $S(k)$ by systematically changing $V_2$ across various electric field cuts (Fig.~\ref{fig:SF-V1-4-h-Combined}) yields two key observations. 
First, at low fields ($h=0$), $S(k)$ clearly traces the expected MI $\to$ LL $\to$ COI progression: the MI phase is defined by a single $k=\pi$ peak, the LL phase shows the $k=\pi$ peak diminishing as $k=\pi/2, 3\pi/2$ side peaks emerge, and the COI phase is dominated by the two side peaks. 
Crucially, as the electric field $h$ increases, the distinction between the LL and COI profiles diminishes rapidly. By $h=4.0$ (bottom panel), $S(k)$ across both regions shows almost exclusively the two prominent $k=\pi/2$ and $k=3\pi/2$ peaks. 
This convergence demonstrates that $S(k)$ alone is insufficient to serve as a definitive criterion for distinguishing the metallic LL regime from the insulating COI at large $h$, which requires the use of the charge gap for comprehensive phase characterization.
%%%%%%%%%%%%%%%%%%%%%%%%%%%%%%%%%%%%%%%%%%%%

\subsection{Pair-pair correlation function results}

The inclusion of the 2NN interactions in the model generally enhances localization effects, favoring short-range correlations and leading to a rapid decay of the pair-pair correlator, $\langle \hat{b}_i^{\dagger} \hat{b}_j \rangle$. 
This observable plays a key role in characterizing the nature of pair formation and coherence in the different phases. Due to the inherent rapid decay of this correlator and the numerical precision limits of our DMRG simulations, we restrict our analysis to values above a practical cutoff of approximately $10^{-20}$, below which the results are indistinguishable from numerical noise.

Figure~\ref{PPC-MI-FL-COI-L128} shows the behavior of the one-length
dimer pair-pair correlator for representative model parameters corresponding to the MI, LL, and COI phases. A clear distinction is observed among the three regimes. In the COI phase (red line), the correlator decays most rapidly, indicating a very short correlation length for pairs. This sharp decay reflects the strong localization of charge carriers and the suppression of long-range coherence. In contrast, the LL (green line) and MI (blue line) phases exhibit a noticeably slower decay, corresponding to more extended correlations.

The pronounced localization observed in the COI phase can be directly attributed to its defining property: pairs become pinned at specific lattice sites forming a periodic pattern, which inhibits their propagation. Consequently, the pair-pair correlator, which measures the probability of finding a pair at site $j$ given one at site $i$, rapidly vanishes as the separation $|i-j|$ increases. This indicates that the order in the COI phase is essentially static and local, with minimal quantum fluctuations allowing pair coherence to extend beyond a few lattice sites. 

In contrast, the LL and MI phases display slower correlation decay, reflecting different microscopic mechanisms:
\begin{itemize}
    \item{the LL phase. 
    Although pairs are formed, they retain a degree of mobility and are subject to quantum fluctuations \cite{Giamarchi2004}. 
    This fluidity allows pair excitations to propagate coherently over larger distances, resulting in a gradual, power-law-like decay of $\langle \hat{b}_i^{\dagger} \hat{b}_j \rangle$.}
    \item{the MI phase. 
    Strong NN intersite repulsion localizes individual particles \cite{Kebric2023} (cf. also \cite{KapciaPRB2017,KapciaPRB2019,AlekseevPRB2025} and reference therein), leading to weak and short-ranged pair correlations. 
    However, unlike in the COI phase, the absence of a rigid charge order pattern permits limited fluctuations, granting slightly more spatial freedom to pair-like excitations.}
\end{itemize}

\begin{figure}[b]
  \centering
    \includegraphics[width=1.0\linewidth,center]{\folderOK 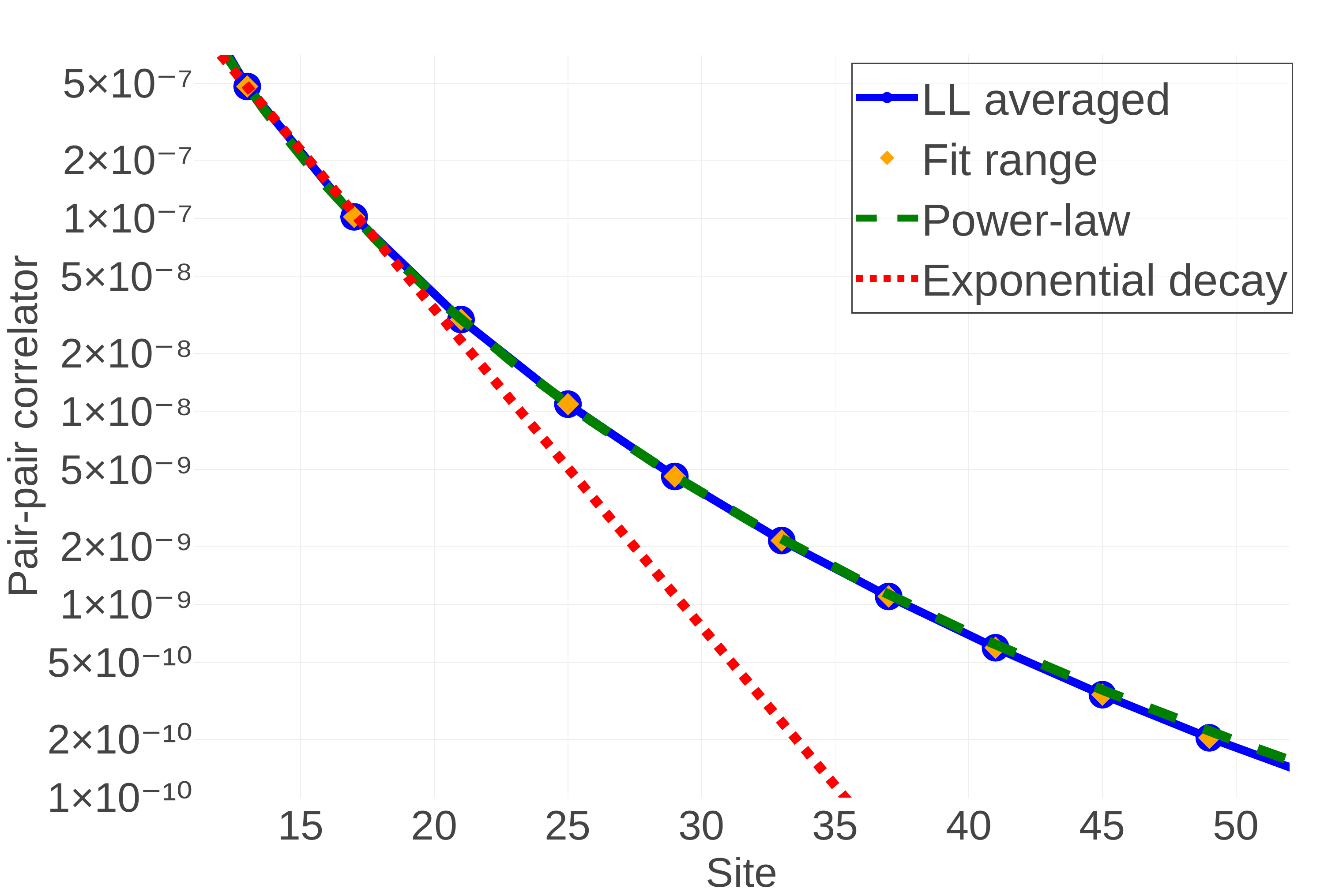}
    \includegraphics[width=1.0\linewidth,center]{\folderOK 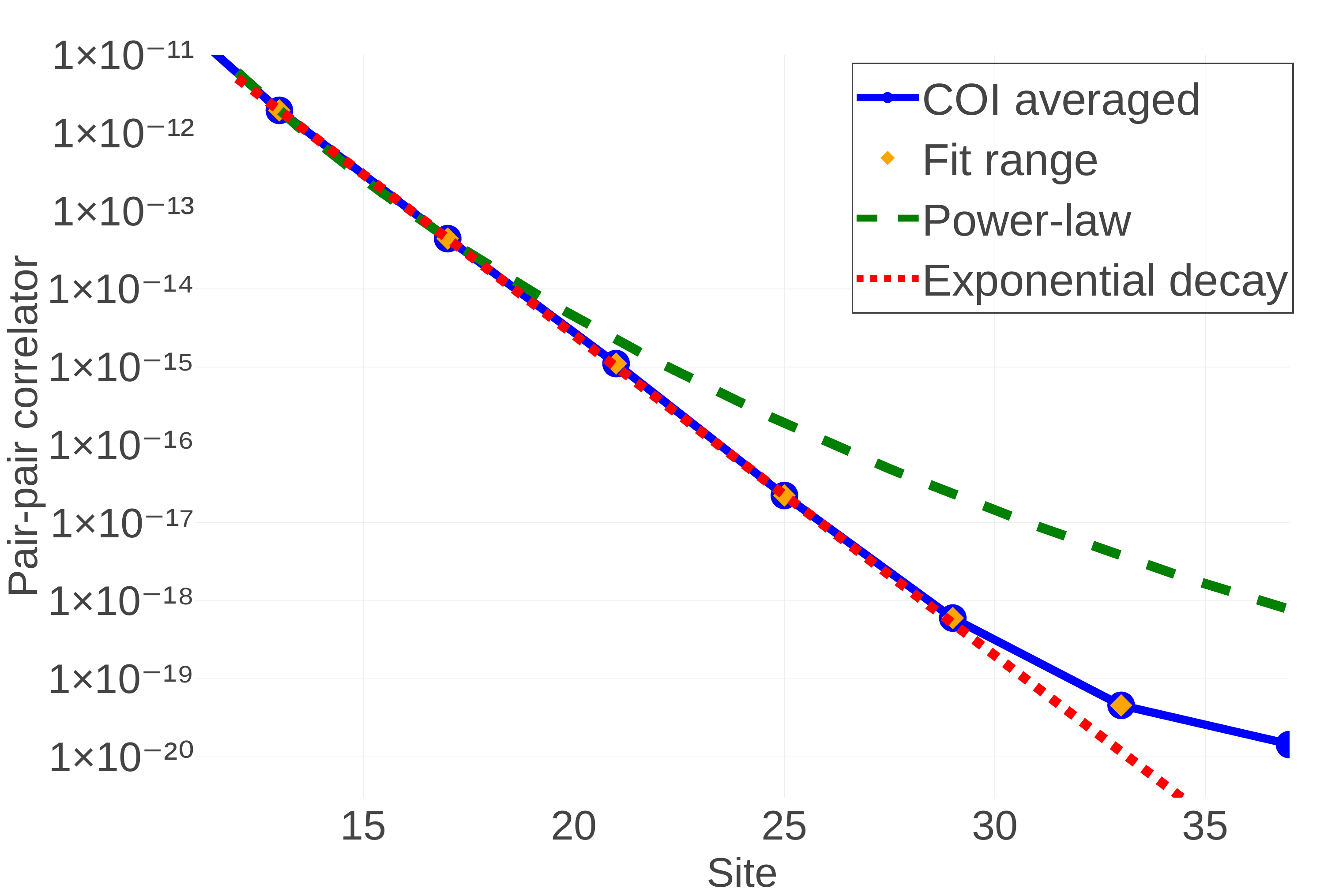}
  \caption{One-length dimer pair-pair correlators for the LL phase ($V_1 = 4.0$, $V_2 = 2.0$, $h = 0.5t$, $L = 128$; upper panel) and the COI phase ($V_1 = 4.0$, $V_2 = 5.0$, $h = 0.5t$, $L = 128$; lower panel). 
  The best fits correspond respectively to a power-law decay, consistent with gapless behavior, and an exponential decay, characteristic of gapped insulating behavior.}
  \label{fig:PPC-FIT-FL-COI}
\end{figure}

To quantify the decay behavior, we fitted the numerical data using two functional forms: a power law, $a x^{-b}$, and an exponential decay, $a e^{-b x}$ ($x$ corresponds to the length, which is proportional to the site distance), see also \cite{BanulsPRD2019,Kebric2023}. Fits were performed in regions where the correlator values remained within numerical precision, and data were averaged over every four sites to reduce fluctuations. The results, presented in Fig.~\ref{fig:PPC-FIT-FL-COI}, demonstrate that the phase transitions are accompanied by a qualitative change in the decay behavior of the correlations. As the system evolves from the LL phase to the COI phase, the decay changes from power-law to exponential, indicating the opening of an excitation gap and the onset of short-range order. 
Conversely, when the system transitions from the MI to the LL phase, the decay changes from exponential to power-law, reflecting the restoration of gapless excitations and long-range coherence. 

This crossover between exponential and algebraic decay is consistent with the general theory of one-dimensional quantum systems, where gapless Luttinger-liquid phases exhibit algebraic correlations, while gapped insulating phases display exponential decay \cite{Giamarchi2004, CalabreseCardy2004}.
This behavior indicates that extended interactions promote localization and charge ordering over coherent pair propagation.

%%%%%%%%%%%%%%%%%%%%%%%%%%%%%%%%%%%%%%%%
\subsection{Entanglement entropy and central charge analysis}

\begin{figure}[b]
    \centering
    \includegraphics[width=1.0\linewidth,center]{\folderOK 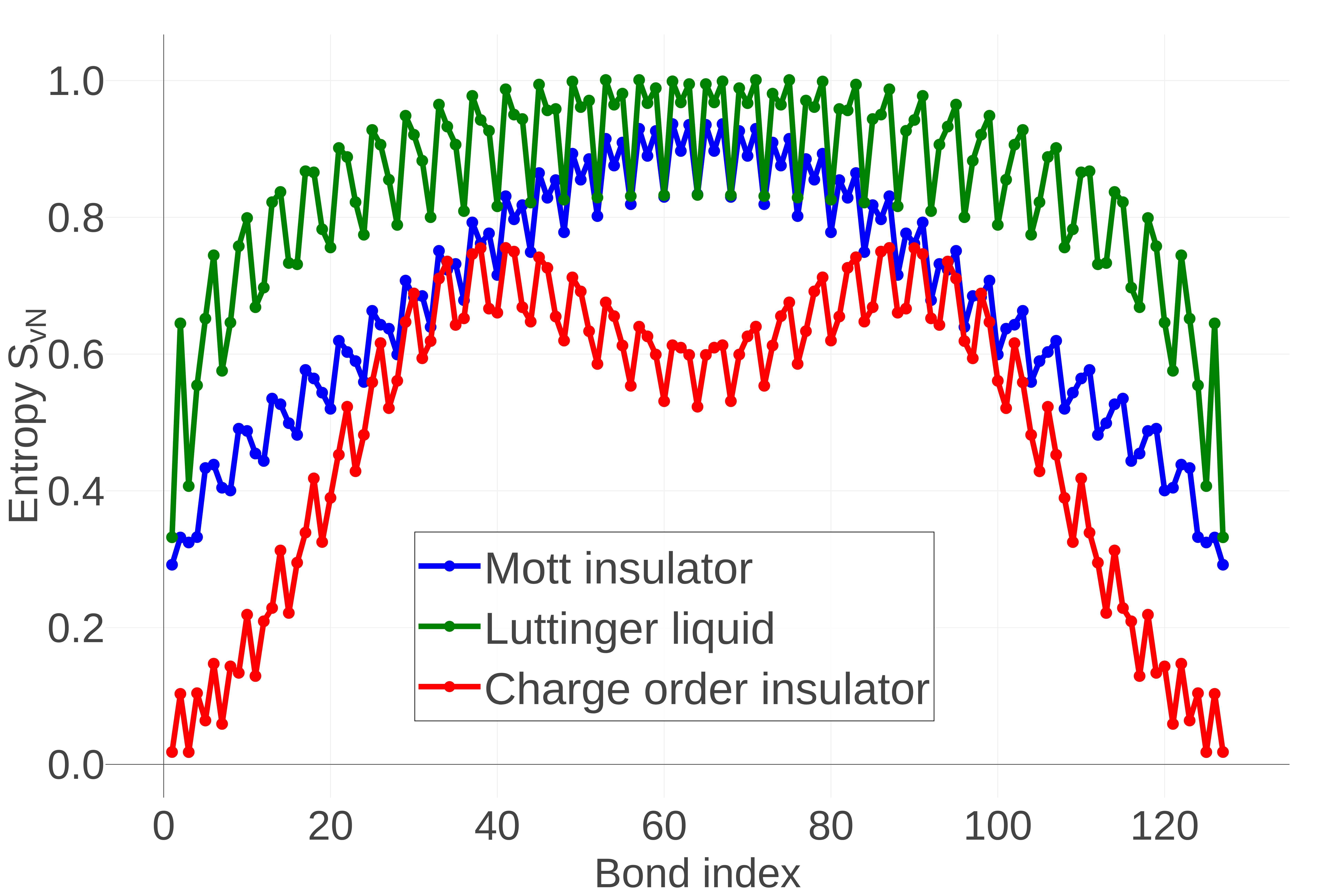}
    \caption{Typical entropy profiles in each phase (as labeled) for $L=128$ and $V_1 = 4.0$: $V_2 = 0.5$ and $h=0.25$ corresponding to MI phase represented by the blue line; $V_2$ = 1.5 and $h=0.5$ in the LL phase (green line); $V_2$=4.5 and $h=1.5$ in the COI phase (red line).}
    \label{Adjusted-Entropy-profiles}
\end{figure}

The spatial profile of the von Neumann entanglement entropy $S_{vN}(b)$ as a function of the bond index $b$ reveals how quantum correlations are distributed along the chain. Distinct characteristic shapes emerge across different phases:
\begin{itemize}
    \item{sharp symmetric central peak indicates short-range correlations typical of gapped or near-critical phases (e.g., the MI phase);}
    \item{symmetric plateau reflects long-range correlations extending over the entire system (e.g., the LL phase);}
    \item{symmetric double-peak structure appears in phases with competing orders or multiple correlation lengths (e.g., the COI phase).}
\end{itemize}
These characteristic profiles act as visual fingerprints of the underlying quantum correlations and, together with other observables, provide a powerful means for phase identification (see Fig.~\ref{Adjusted-Entropy-profiles}).

Additionally, one can observe how entropy is maximum in LL phase and, as we move into gapped phases, $S_{vN}$ is minimized, which is consistent with the fact that in LL phases the particles are mobile, and in COI and MI, there more localized.

Following the discussion in Sec.~\ref{sec:entandcetral}, we determine central charge $c$ by fitting the entanglement entropy profiles, $S_{vN}(b)$, to the functional form of the Calabrese-Cardy relation for $L=128$. This is achieved by plotting $S_{{vN}}(b)$ against $\log\!\left[\left(\frac{2L}{\pi}\right)\sin\!\left(\frac{\pi b}{L}\right)\right]$ and fitting the slope, which yields $c/6$. Values of $c \approx 1$ are used to identify approximate boundaries between gapped (MI or COI) and gapless (LL) phases~\cite{Giamarchi2004, Calabrese2009}.

\begin{figure}[b]
    \includegraphics[width=1.0\linewidth,center]{\folderOK 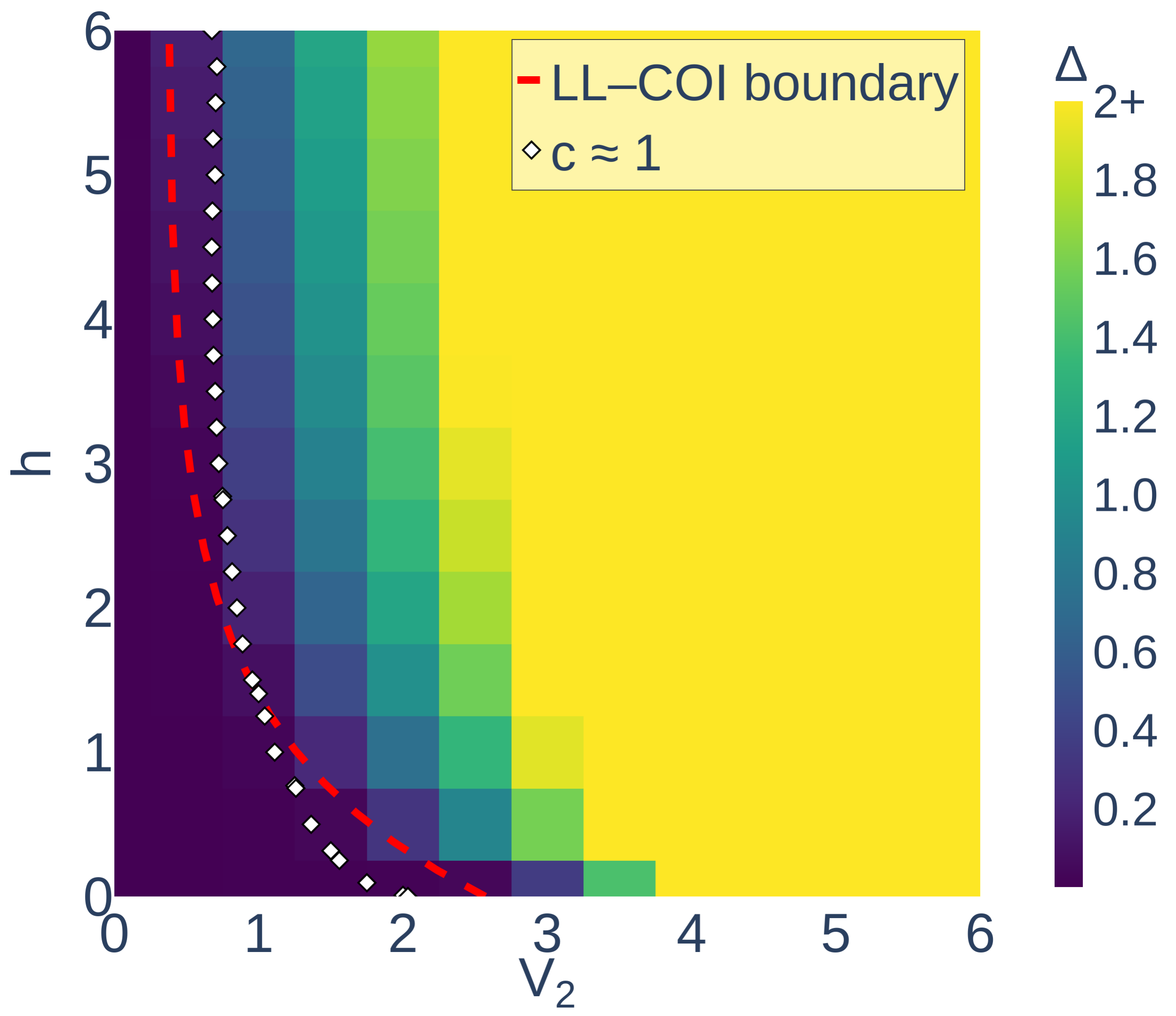}
    \caption{%
    Phase diagram in the ($V_2$, $h$) plane for $V_1 = 1.0$.
    Charge gap $\Delta$ for $L\rightarrow \infty$ is shown as a heatmap and the color bar indicates the magnitude of $\Delta$.  
    The dashed red line represents the numerically determined LL-COI phase boundary. Additionally, the white dots represent the critical boundary points determined by the central charge criterion $c \approx 1$ for a system size of $L=128$.}
    \label{V1_1_phasediagram}
\end{figure}

\begin{figure}[b]
    \includegraphics[width=1.0\linewidth,center]{\folderOK 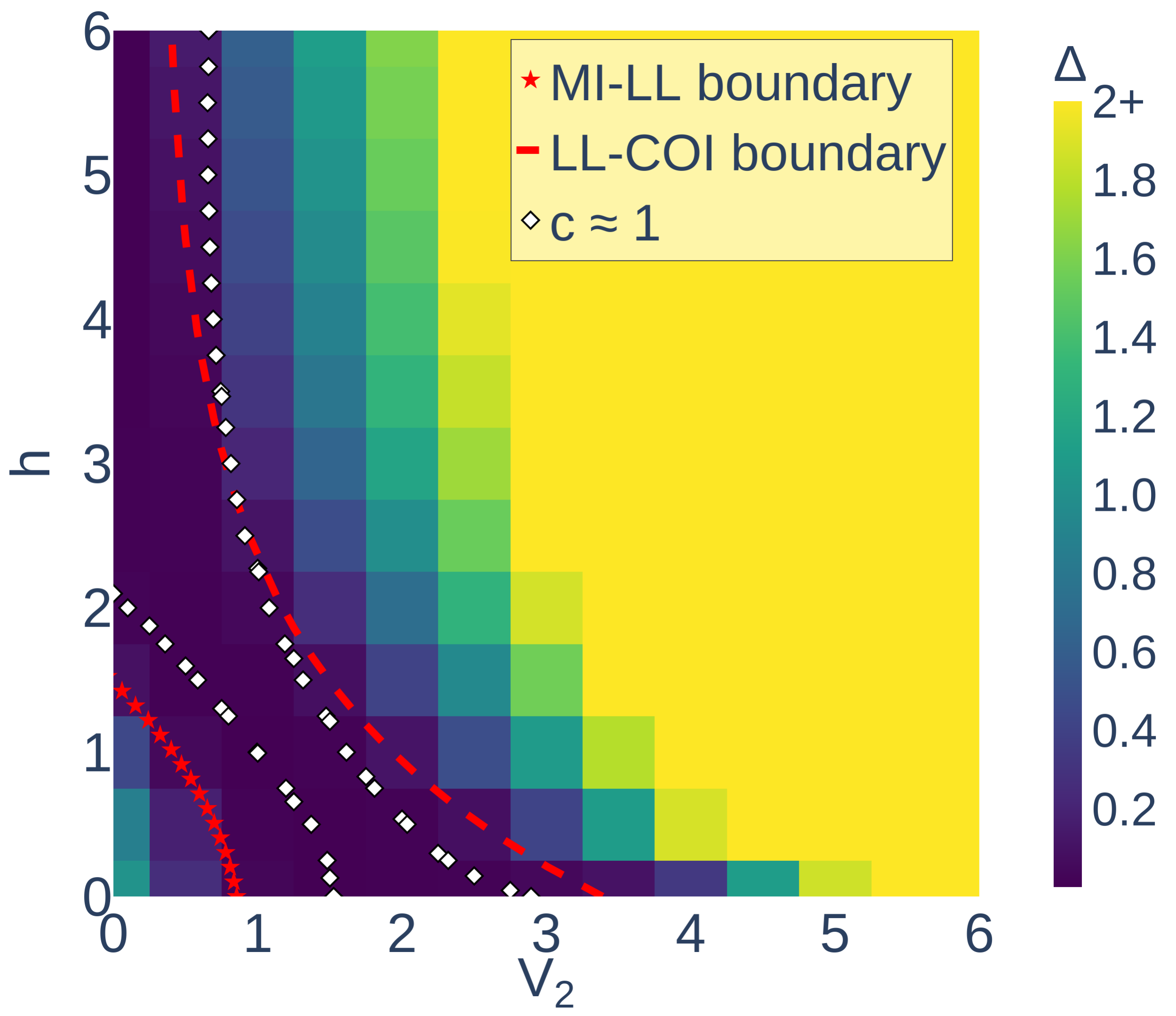}
    \caption{% 
    Phase diagram in the ($V_2$, $h$) plane for $V_1 = 4.0$.
    Charge gap $\Delta$ for $L\rightarrow \infty$ is shown as a heatmap and the color bar indicates the magnitude of $\Delta$. 
    The dotted red line indicates the MI-LL boundary, while the dashed red line marks the LL-COI boundary. 
    Additionally, the white dots represent the critical boundary points determined by the central charge criterion $c \approx 1$ for a system size of $L=128$.}
    \label{V1_4_phasediagram}
\end{figure}

The phase boundaries determined using this $c \approx 1$ criterion are displayed as white dots in Figs.~\ref{V1_1_phasediagram} and \ref{V1_4_phasediagram}, contrasting with the rigorously extracted phase boundaries ($L\rightarrow\infty$) calculated by the charge gap approach $\Delta$ (explained in Section \ref{critical-charge-gap-S}).
It is clear that there is no quantitative agreement between the two estimates due to pronounced finite-size effects in the $L=128$ central charge analysis. 
However, we observe reasonable qualitative consistency, particularly for the LL--COI boundary, where $c\approx1$ is a proxy for the onset of characteristic double-peak behavior in $S_{{vN}}(b)$, which manifests itself deep in the COI phase. 
The $c \approx1$ criterion for the MI--LL boundary, in turn, systematically overestimates the critical values of $(h,V_2)$ determined from the analysis of critical charge gap $\Delta$ (cf. also Sec. \ref{Phase-Diagram-S}). 
Overall, the extracted $c$ values provide qualitative insight into distinguishing gapless versus gapped regions, but are not sufficient for precise phase boundary determination at finite $L$ and serve as complementary evidence for overall phase characterization \cite{Giamarchi2004}.

Note that, while the charge gap provides a direct measure of the metal-insulator transition, central charge is sensitive to logarithmic corrections and finite-size scaling effects that can shift its maximum away from the thermodynamic critical point. 
This behavior is generic in models with complex phase diagrams, such as the present $\mathbb{Z}_2$ LGT with $V_2$ interactions, where the presence of multicritical points or crossover regimes introduces competing correlation lengths. 
Consequently, we treat the charge gap as the primary quantitative tool for boundary determination, while the central charge serves as a qualitative verification of the conformal field theory.

\subsection{Phase diagrams}
\label{Phase-Diagram-S}

\begin{figure}[t]
    \includegraphics[width=1.0\linewidth,center]{\folderOK 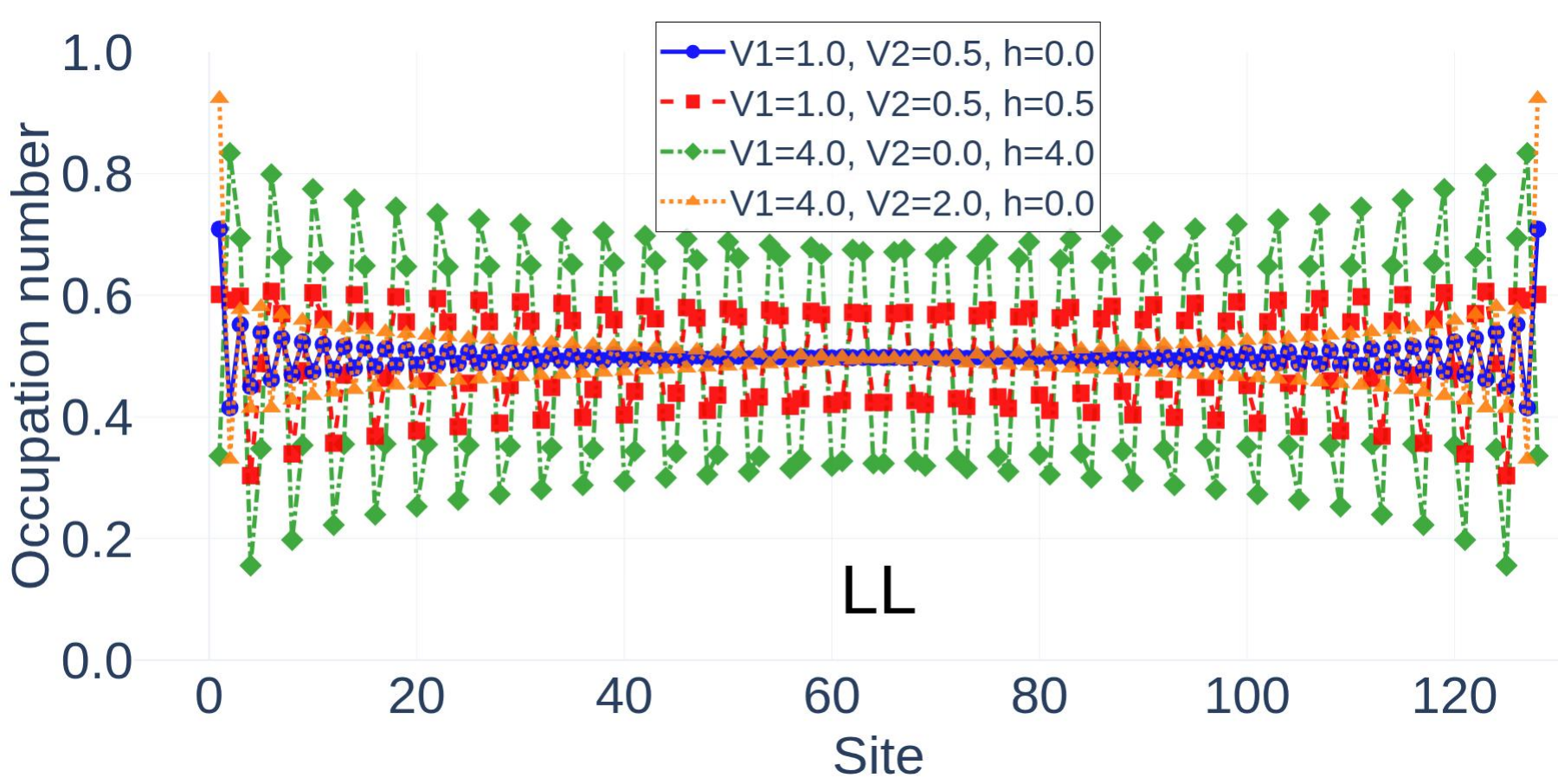}
    \includegraphics[width=1.0\linewidth,center]{\folderOK 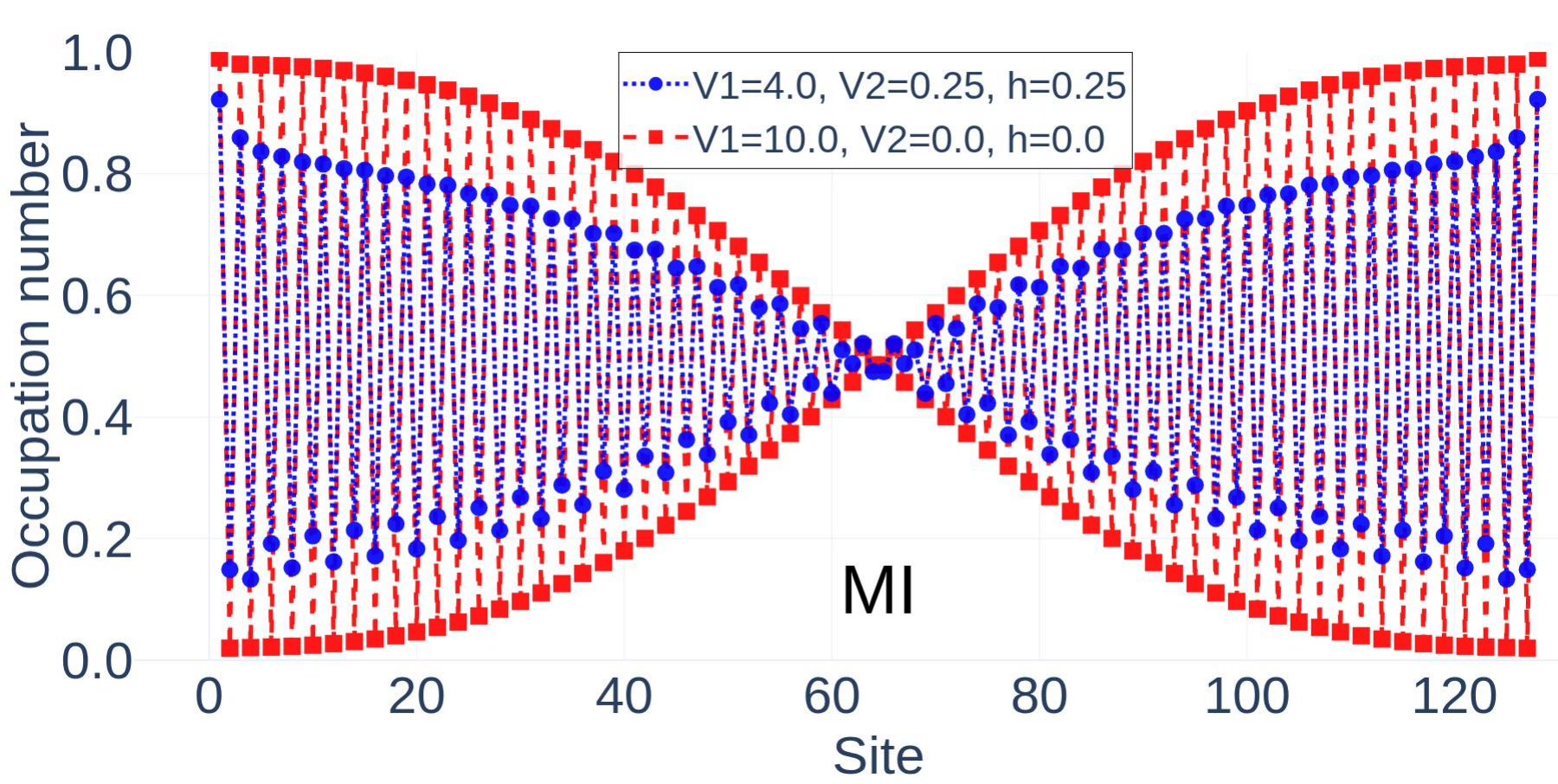}
    \includegraphics[width=1.0\linewidth,center]{\folderOK 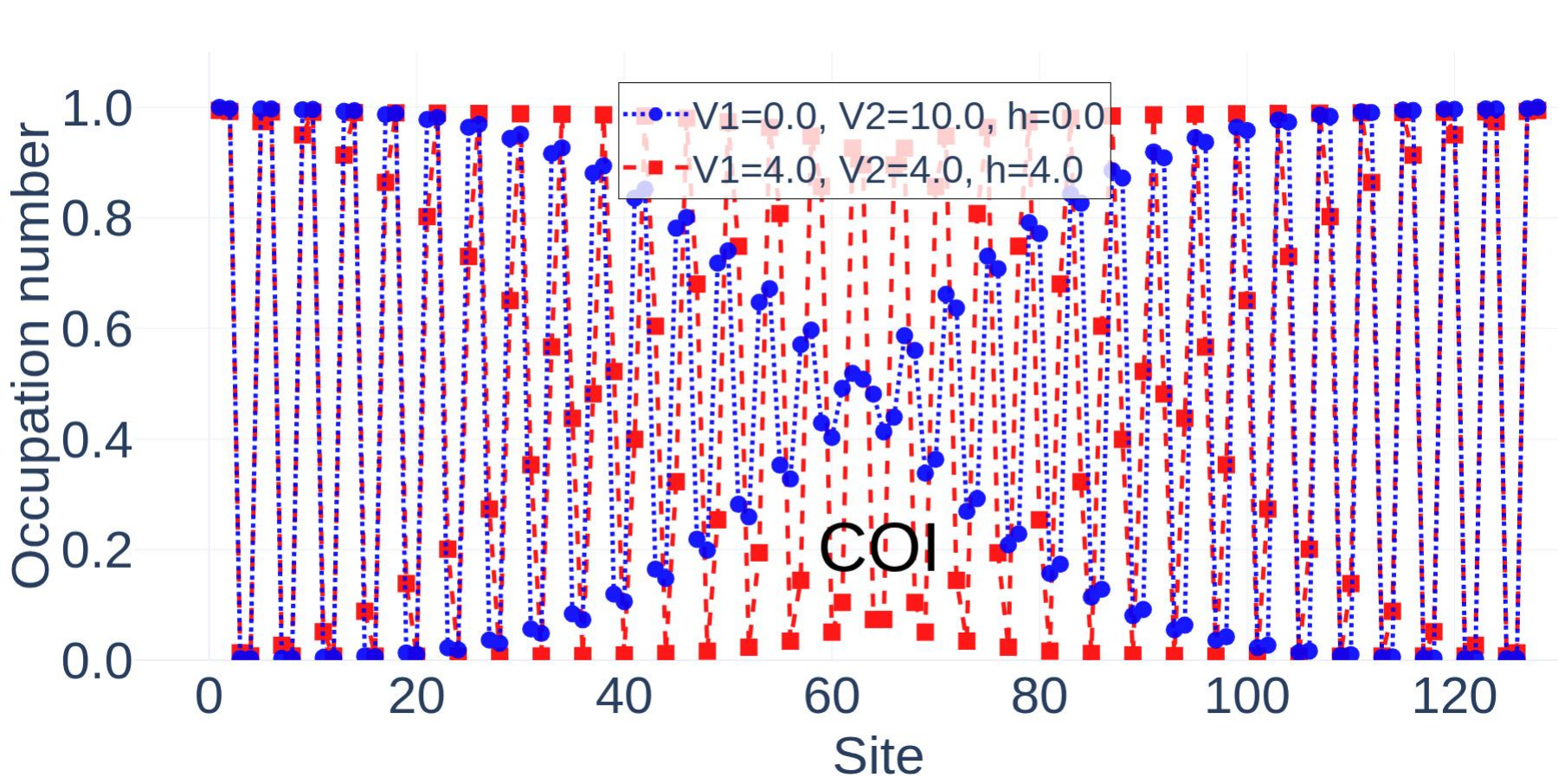}
    \caption{%
    Occupation number as a function of a lattice site for $L=128$ in three phases found: the LL phase, the MI phase, and the COI phase (from the top to the bottom; model parameters as labeled).}
    \label{fig:occupation}
\end{figure}

We summarize our findings by presenting our final phase diagrams, where boundaries between gapped and gapless phases are rigorously determined by identifying critical points where the charge gap vanishes ($\Delta \to 0$) in the thermodynamic limit. 
We classify the  LL phase as gapless ($\Delta = 0$) and the MI and COI phases as gapped ($\Delta > 0$). 
Precise boundaries are estimated by linearly extrapolating $\Delta$ to zero as a function of $V_2$ and $h$ (see Figs. \ref{delta-critical-v1=1} and \ref{delta-critical-v1=4}).
However, since the charge gap provides no distinction between the MI and COI phases, a comprehensive characterization requires complementary observables. 
Therefore, the phase structure is fully validated using a suite of methods: the static structure factor and pair-pair correlation functions confirm the order types of the gapped phases, and the entanglement entropy as well as the resulting central charge serve to reinforce the boundaries and characterization of the three phases. 
Moreover, Fig. \ref{fig:occupation} presents the occupation number as a function of a lattice site in three different phases found: the LL, MI, and COI phases. As one can see from the presented plots, the COI phase exhibits a four-site ordering pattern (AABB), whereas the MI phase shows a two-site ordering pattern (AB, where A are charge-rich sites and B are charge-poor sites). 
In the LL phase, depending on the model parameters both types of arrangement are identified, which are associated with meson-LL and parton-plasma LL states, respectively  (see the discussion at the end of the section).

Our calculations reveal two types of phase diagrams that can be realized in the 2NN model, rigorously constructed based on the $\Delta \to 0$ criterion and supplemented by complementary observables. 
The complete phase diagrams are presented in Figs. \ref{V1_1_phasediagram} and \ref{V1_4_phasediagram}.

For the first case, $V_1=1.0$ (Fig.~\ref{V1_1_phasediagram}), where the 1NN model evinces only the LL phase \cite{Kebric2023}, we observe the emergence of a new COI phase, induced by the 2NN interaction. Thus, the inclusion of $V_2$ enhances confinement. 
The transition from the low-gap (LL) region to the high-gap (COI) region is visible on the diagram.
Along the $h=0$ line, the gap opens at a critical $V_{2c} \approx 2.6t$. The LL-COI phase boundary shown by the dashed red line is qualitatively confirmed by our entanglement entropy and central charge calculations (white dots), which validate the metallic nature of the adjacent LL region.

For the second case, $V_1=4.0$ (Fig.~\ref{V1_4_phasediagram}), the system displays three distinct phases: MI, LL, and COI. In agreement with Ref.~\cite{Kebric2023}, the MI and LL phases already appear in the 1NN model, with $V_1$ driving the transition to the Mott insulating state. The addition of the 2NN interaction $V_2$ further stabilizes the COI phase, which emerges at larger values of $V_2$. 
Accordingly, $V_1$ primarily induces the MI region, while $V_2$ is responsible for the onset of COI. 
The system undergoes a transition from the MI to the COI phase, passing through the intermediate LL phase. 
The figure shows two transition boundaries: the MI-LL (dotted red line) and LL-COI (dashed red line). 
The phase distinctions are most clearly evidenced in the structure factor and pair-pair correlator, which robustly capture the onset of insulating behavior, while the central charge provides additional qualitative support for the identification of the intermediate LL phase.
Note that the greatest difference with respect to the thermodynamic limit is for the MI-LL boundary.
This combined analysis highlights the interplay between $V_1$, $V_2$, and $h$ in shaping the rich phase structure of the system.

Before concluding, it is instructive to comment on the pre-formed parton plasma identified in~\cite{Kebric2023}. 
This regime appears as an extended crossover region separating the MI phase (with localized particles which might be interpreted as partons) from the LL phase (with itinerant particles exhibiting similar density distribution on the lattice as in the MI phase, interpreted as parton-plasma, cf. Fig. \ref{fig:occupation}, the top panel, blue and orange lines).
In the model under consideration, the introduction of a second-nearest-neighbor interaction $V_{2}$ stabilizes a robust COI phase (cf.  Fig. \ref{fig:occupation}, the bottom panel, particularly blue line, deep inside the COI region) and favors the LL phase with four-site-like density distribution (cf. Fig. \ref{fig:occupation}, the top panel, red and green lines, interpreted as a meson-LL state).
The $V_2$ repulsion leads to the stabilization of the COI phase, which is a result of localization of particle pairs in the meson-LL state.

\begin{table*}[t]
\caption{\label{tab:phase_summary}%
Summary of phases in our model including 2NN.
Key observables for each phase: $\Delta$ (critical charge gap), $S(k)$~(static structure factor),
$\langle b_i^{\dagger} b_j \rangle$ (pair-pair correlator), and $S_{vN}(b)$ (entanglement entropy).%
}
\centering
\renewcommand{\arraystretch}{1.3} % Adjust row height
\setlength{\tabcolsep}{2pt}       % Column spacing
\begin{tabular}{c|c|c|c|c}
\toprule
Phase & \boldmath$\Delta $ & \textbf{$S(k)$} & \textbf{$\langle b_i^{\dagger} b_j \rangle$} & \textbf{$S_{vN}(b)$} \\
\midrule
MI &
$\Delta  > 0$ &
max. at $ n\pi$ &
exponential decay &
symmetric sharp central peak \\
LL &
$\Delta  = 0$ &
max. at $(2n+1)\frac{\pi}{2}$ and/or broad max. at $\pi$ &
slow power-law decay &
symmetric broad  plateau \\
COI &
$\Delta  > 0$ &
strong max.\ at $(2n+1)\frac{\pi}{2}$ &
fastest exponential decay &
symmetric with two peaks \\
\bottomrule
\end{tabular}
\end{table*}

\section{Conclusions}
\label{conclusions-S}

This work significantly advances our understanding of the one-dimensional $\mathbb{Z}_2$ lattice gauge theory model at the half-filling by systematically investigating the impact of second nearest-neighbor interactions on its ground-state phase diagram (Table \ref{tab:phase_summary} summarizes our findings). 
Employing matrix product state and density matrix renormalization group techniques, we have calculated key observables: the charge gap, static structure factor, entanglement entropy, and pair-pair correlation functions revealing a far richer and more intricate phase structure than previously understood 1NN model, particularly due to the presence of extended interactions \cite{Defenu2023}.

Our findings highlight distinct behaviors across two representative cases of nearest-neighbor coupling, $V_1$. 
For $V_1 = 1.0$, where the system exhibits a Luttinger liquid phase (for $V_2=0$), the introduction of 2NN interactions drives a direct transition to a charge-ordered insulator phase, demonstrating that $V_2$ interactions play a crucial role in enhancing confinement and restricting the LL region. 
For $V_1 = 4.0$, which corresponds to a Mott insulating phase (for $V_2=0$ and small field $h$), we cover a complex intermediate LL phase that mediates the transition from MI to COI. This emergence of the LL region shows how specific combinations of 2NN interaction ($V_2$) and magnetic field ($h$) can effectively deconfine the system, opening new avenues for controlling quantum phases.

Furthermore, our analysis of the static structure factor ($S(k)$) provides clear signatures distinguishing the MI phase, while also revealing the limitations of $S(k)$ alone in differentiating between the LL and COI phases, particularly at higher field strengths. 
This emphasizes the necessity of complementary observables, such as the charge gap and the entanglement entropy calculations, for a robust phase characterization. 
The study of pair-pair correlators further corroborates the role of 2NN interactions in promoting short-range correlations and suppressing long-range pair coherence, notably in the COI phase, where pair propagation is significantly limited.

The entanglement entropy profiles $S_{vN} (b)$ provided a robust characterization of all phases identified in the phase diagram. We observed that the profiles for both the gapless LL and the gapped MI phases are symmetrical with respect to the half-chain cut ($L/2$). However, their shapes differ critically: the LL phase exhibits a wide, symmetric plateau near $L/2$, a signature consistent with the dominance long-range fluctuations. In contrast, the MI profile displays a sharper, localized peak at $L/2$, reflecting the short correlation length of the gapped state. 
The most distinctive result is found in the COI phase. 
Here, the profile retains symmetry but develops two pronounced peaks located away from the center (at $L_A \approx L/4$ and $3L/4$), separated by a central minimum at $L_A=L/2$. This unique double-peak structure is a clear hallmark of a long-range periodic order, where the entanglement is minimized when the partition boundary coincides with a nodal point.

The gapless phases and quantum critical points in the 1D $\mathbb{Z}_{2}$ LGT are fundamentally expected to be described by conformal field theories, making the central charge ($c$) a standard tool for phase identification. 
However, we find compelling evidence that, for the complex phase structure of the 1D $\mathbb{Z}_{2}$ LGT model with 2NN interactions, the traditional classification based on the central charge extracted from finite-size scaling of the entanglement entropy can yield apparent phase boundaries that deviate from the true thermodynamic transitions, due to finite-size effects. 
Our analysis demonstrates that the charge gap ($\Delta$) serves as a more robust and definitive criterion for identifying the quantum phase boundaries in the thermodynamic limit, highlighting the limitations of finite-size $c$ analysis near multicritical and/or crossover regimes in such systems.

In the present study, the COI phase is stabilized by the second-nearest-neighbor interaction $V_2$ at half-filling. This phase is characterized by a four-site unit cell ordering pattern (schematically illustrated as $\ldots 11 00 1100 \ldots $ in the occupation basis, where $1$ and $0$ correspond to a site with high and low particle concentration, respectively). 
Adding longer-range interaction can change the phase diagram and introduce new phases. In particular, one can expect that adding a third-nearest-neighbor (3NN) term $V_3$ might stabilize a new ordered state, with six-site unit cell pattern $\ldots 111 000 111 000 \ldots $ for $V_3 \gg V_1, V_2, t$. Detailed analysis of the properties of this new phase is out of the scope of the present work and we leave it to forthcoming works.

It is worth emphasizing that the phase structure uncovered in the one-dimensional $\mathbb{Z}_2$ lattice gauge theory with 2NN interactions can be probed directly in state-of-the-art quantum-simulation platforms capable of resolving gauge-invariant degrees of freedom with high fidelity. 
In ultracold atom realizations based on optical lattices or Rydberg-dressed ensembles, gauge-invariant observables such as string operators, Wilson-line correlators, or electric-field expectation values can be reconstructed from single-site and single-link readout using quantum gas microscopy, providing a direct route to distinguishing confined, deconfined, and symmetry-broken regimes \cite{Wiese2013,Goldman2016}. 
The inclusion of second–neighbor couplings generically enhances the sensitivity of parity-resolved density correlations and allows the identification of translation‑symmetry breaking through the appearance of momentum‑space peaks in the static structure factor $S(k)$, which in turn can be extracted via time‑of‑flight imaging or Bragg spectroscopy \cite{Kuhr2016}. 
Furthermore, dynamical probes, such as quench spectroscopy across the phase boundaries, yield characteristic signatures in the real‑time evolution of electric‑field strings and domain‑wall densities. 
In particular, the relaxation dynamics of non‑equilibrium strings, already observed in Rydberg‑atom experiments simulating $\mathbb{Z}_2$ gauge theories \cite{Bernien2017}, provide a sensitive probe of the underlying phase and may reveal the modified confinement scales induced by second‑neighbor interactions. 
Trapped‑ion architectures offer complementary access routes: the encoding of gauge and matter fields in internal spin states and phonon‑mediated interactions enables direct measurement of $\tau^x$ and $\sigma_z$ correlators, and recent experiments have demonstrated the capability to track emergent gauge‑field dynamics in real time with high resolution \cite{Zhang2017,Monroe2021}.

In conclusion, the convergence of Rydberg‑atom arrays for simulating $\mathbb{Z}_2$ gauge theories \cite{Surace2020} ensures that the phase structure reported in this work is within the reach of current experimental resolution. 
These platforms provide a robust framework for investigating the stability of the COI phase and the modified confinement scales arising from second‑neighbor couplings.

In summary, this work not only extends the known phase landscape of 1D $\mathbb{Z}_2$ LGT models but also provides fundamental insights into the intricate interplay between gauge fields, confinement mechanisms, and extended interactions.

%%%%%%%%%%%%%%%%

%

\end{document}